\RequirePackage[log]{snapshot}
\documentclass[11pt]{article}
\usepackage{graphicx,amssymb,amsfonts,amsmath,subfigure,a4,authblk}
\usepackage[round]{natbib}
\newcommand*\mylabel[1]{\label{#1}}

\newcommand{\co}{conditional }

\newcommand{\G}{GARCH }

\newcommand*\myref[1]{(\ref{#1})}

\begin{document}
\title{MULTIVARIATE GARCH WITH DYNAMIC BETA}
\author[1,2]{M. Raddant}
\author[3]{F. Wagner}
\affil[1]{Institute for the World Economy\\
Kiellinie 66, Kiel 24105, Germany}
\affil[2]{Department of Economics, Kiel University\\
Olshausenstr. 40, 24118 Kiel, Germany\\
raddant@economics.uni-kiel.de} 
\affil[3]{Institute for Theoretical Physics, Kiel University\\
Leibnizstr. 19, Kiel 24098, Germany, 
wagner@theo-physik.uni-kiel.de}
\maketitle

\begin{abstract}
We investigate a solution for the problems related to the application of multivariate 
GARCH models to markets with a large number of stocks by restricting the
form of the \co covariance matrix. The model is a factor model and uses only six free GARCH parameters. One factor can be interpreted
as the market component, the remaining factors are equal. This allow the analytical calculation of the
inverse covariance matrix. The time-dependence of the factors enables the determination
of dynamical beta coefficients. We compare the results from our model with the results of other GARCH 
models for the daily returns from the S\&P500 market and find that they are competitive. As applications we use the daily 
values of beta coefficients to confirm a transition of the market in 2006.
Furthermore we discuss the relationship of our model with the leverage effect.
\end{abstract}

\section{Introduction  \label{intro}}
The estimation of co-movement between stocks is an essential problem in
the analysis of asset returns, financial integration, and for portfolio management.
On the one side, the statistical properties of asset returns
necessitate to treat them in a setting that includes
time-varying volatility.
The most common setting thus are GARCH models.
In many analysis of co-movement one would however also like to analyze samples that cover entire markets and which can thus be very large.
This is at odds with many multivariate versions of GARCH models.

A multivariate GARCH model for $N$ stocks can be characterized by the dynamics of the conditional
covariance matrix $H(t)$ and by its mean $\bar{H}$. The number of so-called
nuisance parameters in $\bar{H}$ typically exceeds the power of a maximum likelihood 
estimate (MLE) for samples like the constituents of the S\&P index. A possible solution are methods using covariance targeting \citep{EngleMez} that replace $\bar{H}$ by the observed time-averaged covariance $C$. Since $C$ has similarity with a random matrix one can expect the uncertainty in $\bar{H}$ to be in
the order of $\sqrt{N/T}$  \citep[see][]{market,marpa}. 
A different solution are shrinkage methods as proposed by \cite{large_dcc2} which renders improved
estimates of the eigenvalues of $\bar{H}$ but not its eigenvectors. 

We will
overcome the problem of estimating $\bar{H}$ in large samples by a restricted
form of $\bar{H}$ as in the $K$-factor model \citep[see][]{bauwens}. In this method 
only parts of $C$ are used that can be determined with a relative accuracy in the
order of $\sqrt{1/T}$.

Another issue for estimations are the GARCH parameters that govern the dynamics of $H(t)$. 
In the VEC model
\citep{vbeta} the elements of $H(t)$ are treated as a $N(N+1)/2$ vector, which can 
lead to a large number of parameters. In its scalar version \citep{svec} only two 
parameters remain. The $K$-factor BEKK model \citep{en_kron} uses $N$-dimensional 
matrix multiplication. Also in this model the number of parameters can become very large unless
one uses the scalar version with a manageable number of $2K$ parameters.

Last but not least use of MLE requires the estimation of $H^{-1}(t)$ when Gaussian noise is used, and even worse $H^{-1/2}(t)$ for general noise. For large $N$ numerical inversion can be
prohibitive. In $K$-factor
models these time consuming calculations can be done outside MLE resulting in $K$
independent univariate GARCH models for the factors. 
A restricted form  for $H(t)$ has
also been proposed in the DECO model \citep{deco} which assumes the same time-dependent correlation
between all pairs of stocks and therefore also circumvents these issues.


Our model is loosely based on the BEKK model and adds two effective dynamic factors. 
 We define one common volatility factor $v_0(t)$ and
$N-1$ degenerate equal factors $v_1(t)$. This ensures the existence of $H^{-1}(t)$
and at the same time allows its analytical calculation. 
As the second part we use a rotation of
the eigenvector that belongs to the $v_0(t)$ factor. In contrast to the 
GO-GARCH model \citep{GOGarch} this rotation is time-dependent and is determined
dynamically. 
This part of the model therefore delivers
time-dependent beta coefficients relative to the market in the spirit of a CAPM.

Summarized, the model dynamics lead to a coupled system of GARCH recursions for the
factors $v_k(t)$ and a vector recursion for $\beta(t)$. Only six GARCH parameters have to be estimated by MLE. 

The paper is organized in the following way.
The definition of our model and the derivation of the recursion are described in section \ref{multi} together with a comparison to other models. We apply the estimation by MLE
to the daily returns of 356 stocks from the
S\&P market in the years 1995-2013 in section \ref{fits}. We then compare the estimation results with other models in section \ref{data}.
Section \ref{Applications} contains two applications. By using our $\beta(t)$ we verify a transitions in the market observed by \cite{Radd}.
As a second application we investigate the correlation of the
leverage effect with $\beta$.  The last section contains some conclusions.

\section{Multivariate GARCH with restricted covariance}\label{multi}
\subsection{ Model for the covariance matrix}

We consider a time series of returns for $N$ stocks with length $T$ denoted by $r_i(t)$
with $i=1\dots N$ and $t=1\dots T$. We assume the mean of $r_i(t)$ with respect to $t$ 
is zero. In a multivariate GARCH model $r(t)$ are related to a noise $\varepsilon(t)$ by

\begin{equation} \mylabel{a0}
r(t)={H(t)}^{1/2} \; \varepsilon(t)
\end{equation}

\noindent 
The i.i.d. $\varepsilon$ have mean zero  and variance one. The matrix $H(t)$ corresponds to
 the conditional covariance of $r_i(t)$.
The dynamics of
$H(t)$ are expressed by a recursion formula that relates $H(t+1)$ to the values of $H(t)$
and the returns at time $t$. A well-known example for such a model is the $K$-component BEKK(1,1,K) model \citep{en_kron} in which the recursion is normally stated like this
 \begin{equation}\label{eq:rec}
 H(t+1)=\Omega + \sum_k A_k  \bigl( r(t) r(t)' \bigr)  A_k' +
          B_k  H(t)  B'_k 
\end{equation}

\noindent 
The model that we are going to introduce is different from the BEKK model, however, the structural form of the recursion for $H(t)$ has some similarities. As a starting point, consider the following specification

\begin{equation}\label{eq:bekk}
H(t+1)=H(t)+\sum_{k=0}^{K-1} \bigl[ A_k  \bigl(r(t) r(t)'-H(t) \bigr) A_k  
      + G_k \bigl( \bar{H} -H(t) \bigr) G_k \bigr]
\end{equation}

\noindent 
where $A_k$ and $G_k$ are time-independent symmetric $N\times N$ matrices.
$\bar{H}$ denotes the expected value of $H(t)$.\footnote{Note that we use $(t+1)$ as the one-step-ahead index. Note also that (\ref{eq:bekk}) only becomes a properly defined process for the covariance matrix once we define the process for the market volatility, which indirectly imposes restrictions on $A$, $G$ and $H(0)$ and leads to the recursion in (\ref{R}), see also appendices \ref{recurs} and \ref{noise}.}

\noindent
There are several important differences in this equation compared to other GARCH models. 
In many applications of multivariate GARCH models a recursion like equation    
\eqref{eq:bekk}
applies to the standardized returns and  
the correlation matrix $R$. In our case it applies to the conditional covariance matrix $H(t)$. This means that we preserve the information on the level of $r$, which will help to define $\beta$-values.
The recursion also exhibits clearly the fixed
point \mbox{$H(t)=\bar{H}\;$ if $\; r(t) r(t)'$ } is replaced by its conditional expected
value. The parameters in 
$A_k$ describe how fast $H(t)$ returns to its mean $\bar{H}$ after a disturbance by
$r(t) r(t)'$, $G_k$ describes the persistence of $H_t$ (similar yet not identical to $B$ in other GARCH models). This specific form of \eqref{eq:bekk} will be helpful in the following
generalization for time-dependent $A_k,G_k$. 

The linearity of equation \eqref{eq:bekk} also allows the evaluation with
linear algebra.  
We can use this property by using the following representation of symmetric
matrices $M$ with dimension $N$
\begin{equation} \mylabel{a2}
M=\sum_{k=0}^{K-1}\;\lambda_k \;P_k
\end{equation}
with $N\times N$ projection  matrices $P_k$ satisfying the orthogonality relations
\begin{equation} \mylabel{a3}
P_k\cdot P_l=P_k \; \delta_{l k}
\end{equation}
and the completeness relation with the unit matrix $I$
\begin{equation} \mylabel{a4}
\sum_{k=0}^{K-1}\;P_k=I
\end{equation}

It is important to note that the sums in equations \myref{a2} and \myref{a4} run only
over the different eigenvalues $\lambda_k$. The trace of $P_k$ gives the multiplicity of
$\lambda_k$. In the non-degenerate case ($K=N$) the $P_k$ are the dyadic products 
of the eigenvectors. For $K<N$ \; $P_k$ are independent of the particular 
choice of the eigenvectors. Functions of $M$ as for $H(t)$ in equation \eqref{eq:bekk} can 
be calculated with
\begin{equation} \mylabel{a5}
f(M)=\sum_{k=0}^{K-1}\;f(\lambda_k) \;P_k
\end{equation}
This formalism is particular useful in the case of a single eigenvalue
$\lambda_0$ and $N-1$ degenerate eigenvalues $\lambda_1$ ($K=2$). From
equation \myref{a4} we see that essentially only one projector $P_0$ exists
with $P_1=I-P_0$.

\begin{figure}[htb]
\begin{center}
\includegraphics[width=0.70\linewidth, trim = 30 0 30 10, clip=true]{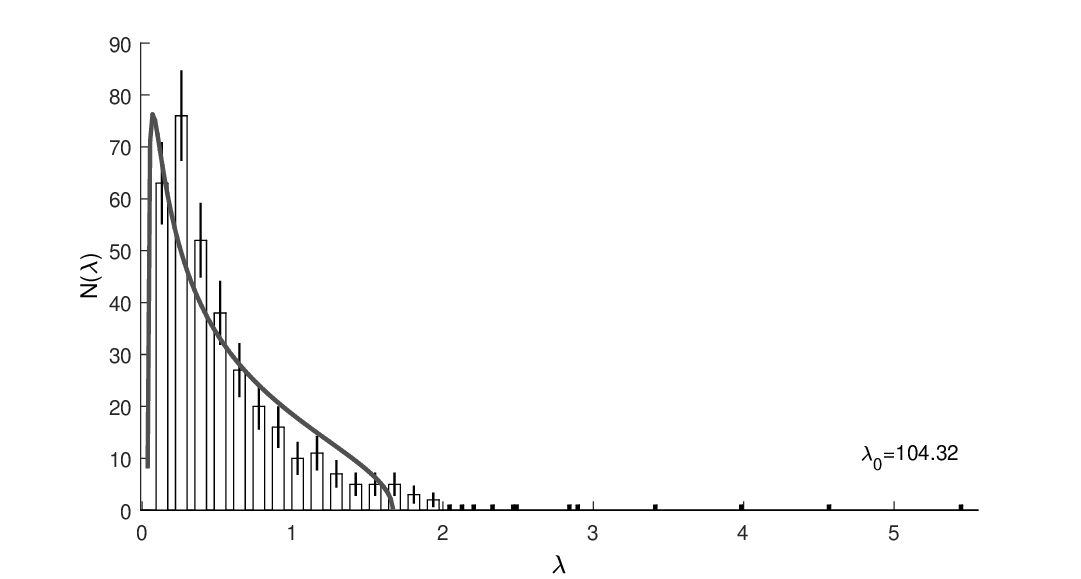}
\end{center}
\caption{\label{c_spectrum} \em Histogram of eigenvalue spectrum of the covariance matrix 
        $C$ for the daily  returns of 356 S\&P constituents. 
        Returns are normalized to have $trace(C)=N$. The largest eigenvalue 
        $\lambda_0=104$ is not shown.  The curve corresponds to a MP spectrum.
        See appendix \ref{sec:chi} for information on the calculation of the errors.}
\end{figure}

The eigenvalue-spectrum of the covariance matrix is related to the projection matrices by
\begin{equation} \mylabel{a6}
C=\frac{1}{T}\sum_{t=1}^T \;r(t) r'(t)=\sum_{k=0}^{N-1}\; \lambda_{k}\; P_k
\end{equation}
In figure \ref{c_spectrum} we show the eigenvalue spectrum for the S\&P data ($\mbox{N=356}, \mbox{T=4782}$).
 $C$ exhibits one 
large eigenvalue $\lambda_0$ and $N-1$ eigenvalues of order $\lambda_0/N$. Within the framework of a multivariate GARCH model one can interpret this behavior as if the true covariance matrix $E[H]$ had a single eigenvalue
$\lambda_0$ and $N-1$ degenerate eigenvalues. The noise transforms them into a MP spectrum as described by \cite{marpa}, which qualitatively agrees with the observed spectrum of $C$.

To arrive at a new form of 
$H(t)$ we assume a two component structure as in $C$ at each $t$ and write $H(t)$ as
\begin{equation} \mylabel{a7}
H(t)=N\;v_0(t)\;P_0(t)\; +v_{1}(t)(I-\;P_0(t))
\end{equation}
with the projector $P_0(t)$
\begin{equation} \mylabel{a8}
P_0(t)=\frac{1}{N}\beta (t) \beta'(t)
\end{equation}
in terms of the eigenvector $\beta$, which is normalized to $\beta'(t) \cdot \beta(t)=N$.

The form assumed in equation \myref{a7} actually corresponds to a $K=N$ factor model. The 
factors are the eigenvalues of $H(t)$. The first factor is $Nv_0(t)$. 
  This component of the large eigenvalue of $C$
can be regarded as the market \citep[see, e.g., ][]{market} and therefore the first part in equation \myref{a7} can be interpreted as a time-dependent market
factor. When we project $r(t)$ on the eigenvector $\beta (t)$ we obtain a market return 
$r_M(t)$ 
\begin{equation} \mylabel{a9}
r_M(t)=\frac{1}{N} \beta'(t)  r(t)
\end{equation}
Due to the relation
\begin{equation} \mylabel{aa9}
\beta_i(t)=\frac{E_{t-1}[r_i(t)r_M(t)]}{E_{t-1}[r_M^2(t)]}
\end{equation}
$\beta_i(t)$ can be interpreted similar to beta coefficients in a CAPM
approach \citep{sharpe,lintner} relative to $r_M(t)$. A similar definition for 
conditional betas can be found in \cite{Engle_DCB}.
Additionally we have $N-1$ degenerate factors $v_1(t)$ which together form the second component and show as the second term in equation \myref{a7}.

We can now apply this two component structure and modify the recursion  into a  factor 
model \citep[see][]{Lin} with time-dependent $A_k$ and $G_k$ with
\begin{equation} \mylabel{a8a}
A_k(t)=\sqrt{\alpha_{k}}\; P_k(t) \quad \mbox{and} \quad  G_k(t)=\sqrt{\gamma_{k}}\; P_k(t)
\end{equation}
In order to make also the $\beta$ in the recursion dynamic we have to add the off-diagonal terms that allow for a transition between 
$P_0(t)$ and $P_1(t)$. These terms contribute only to $r(t) r(t)'$ and $\bar{H}$.
After adding those terms our recursion reads
\begin{align} \mylabel{R}
H(t+1)= H(t) & +   \sum_{k=0,1}P_{k}(t)  \left[\alpha_{k} \left(r(t) r'(t)-H(t)\right)
+\gamma_{k} \left(\bar{H}-H(t)\right)\right] P_{k}(t) \nonumber \\
 & + P_{0}(t) \left[\alpha_{01} r(t) r'(t)+\gamma_{01}\bar{H}\right] P_{1}(t)
\nonumber \\ 
 & + P_{1}(t) \left[\alpha_{01} r(t) r'(t)+\gamma_{01}\bar{H}\right] P_{0}(t)    
\end{align}
The six parameters of our model can be estimated by maximum likelihood, for details see appendix \ref{like}.

For the mean value $\bar{H}$ we assume the same restriction as for $H(t)$
\begin{equation} \mylabel{ab9}
\bar{H}=\bar{v}_0\; \bar{\beta} \bar{\beta}'
+\bar{v}_1(I-\frac{\bar{\beta} \bar{\beta}'}{N})
\end{equation}
Details on covariance targeting and the expected error are described in detail in appendix \ref{noise}. We can  determine the parameters $\bar{v}_{0,1}$ 
 and $\bar{\beta}$ from the observed covariance matrix $C$ with relative accuracy 
$\frac{1}{\sqrt{T}}$ under the assumption of slowly varying $v_0(t)$ and $\beta(t)$.

The recursion for $H(t+1)$ is obviously tied to the recursions for
$v_{0,1}(t+1)$ and $\beta(t+1)$. By
performing the operation \mbox{$tr \;P_k(t) (\cdot)\;$} on both sides of equation \myref{R} one obtains two equations
for $v_k(t+1)$. If we apply both sides of equation \myref{R} on $\beta_i(t)$ we find
a vector $D_i(t)$ which determines the change of $\beta (t+1)$:
\begin{equation} \mylabel{a11}
D_i(t)=\alpha_{01}\; r_M(t)\big(r_i(t)-r_M(t)\beta_i(t)\big) +\gamma_{01}\bar{m}(t)
            \bar{v}_0\big(\bar{\beta}_i-\bar{m}(t)\beta_i(t)\big)
\end{equation}
The details are given in appendix \ref{recurs}. Since the results look rather 
complicated, we quote them here only in the limit $N \gg 1$. In the application
to the S\&P market we found no difference in using the full solution from 
appendix \ref{recurs}.  The recursion for $v_0(t+1)$ is given by
\begin{equation} \mylabel{a10}
m^2(t)v_{0}(t+1)=R_0(t)
\end{equation}
with $R_0(t)$ denoting the operation $tr \;P_0(t)(\cdot)$ on the r.h.s. of equation \myref{R}
\begin{equation} \mylabel{a10a}
R_0(t)=v_{0}(t)+\gamma_{0}\big( \bar{m}^2(t)\bar{v}_{0}-v_{0}(t)\big)+ 
             \alpha_{0}\left(r_M^2(t)-v_{0}(t)\right)
\end{equation}

Equation \myref{a10} depends on the angle $\phi(t)$ between $\beta(t+1)$ and $\beta(t)$ 
given by the overlap $\cos(\phi(t))={m}(t)=\beta'(t+1)\cdot \beta(t)/N$. Similarly, 
$\bar{m}(t)=\bar{\beta}'\cdot \beta(t)/N$ corresponds to the angle between $\beta(t)$ 
and $\bar{\beta}$. 
The overlap ${m}(t)$ is determined from the normalization condition 
$\beta'(t+1)\cdot \beta(t+1)=N$ with
\begin{equation} \mylabel{a12}
m^2(t)=\left[1+\frac{D'(t)\cdot D(t)}{N\;R_0^2(t)} \right]^{-1}
\end{equation}
The recursion for $\beta (t+1)$ is expressed in terms of $D(t)$ by
\begin{equation} \mylabel{a13}
\beta_i(t+1)={m}(t)\left[ \beta_i(t)+\frac{D_i(t)}{R_0(t)} \right]
\end{equation}
$\beta_i(t+1)$ can change only for deviations of $r(t)$
from the market return $r_M(t)$ and deviations of $\beta(t)$ from the mean value $\bar{\beta}$.
Since $D'(t)\cdot D(t) \propto N$ 
there is no $N$-dependence in equation \myref{a12}. When $v_{0}(t+1)$ is known we can obtain the recursion for $v_{1}(t+1)$:
\begin{align} \mylabel{a14}
  v_{1}(t+1) = & \; v_{1}(t)-(1-m^2(t))v_{0}(t+1)
+ \alpha_{1}\bigl(\frac{r^2(t)}{N}-r_M^2(t)-v_{1}(t)\bigr) \nonumber \\
 & +\gamma_{1}(\bar{v}_{1}+
        (1-\bar{m}^2(t))\bar{v}_{0}-v_{1}(t)) 
\end{align}
The returns $r(t)$ appear in the recursions \myref{a10} through the market component $r_M(t)$
and in \myref{a11} and \myref{a14} through the component perpendicular to $\beta(t)$.
The recursions differ from those in factor
models with constant eigenvectors of $H$ 
in the interpretation of $\bar{v}_{k}$ and $\bar{\beta}$. They are determined by
covariance targeting from the empirical covariance matrix $C$ (see appendix \ref{noise}).
Due to the non-linearity of the recursions they may differ from the expectation
values of $v_k(t)$, resp. $\beta(t)$. For small $\alpha_{01}$ the recursions have
a fixed point solution $v_{k}(t)=\bar{v}_{k}$ and $\beta(t)=\bar{\beta}$.
A necessary condition for its stability  are the following
inequalities for the GARCH parameters $\alpha_{k}$ and $\gamma_{k}$
\begin{equation} \mylabel{a15}
0<\gamma_{k}<\gamma_{k}+\alpha_{k}<1
\end{equation}

\subsection{Relation to other models}\label{sec:models}

One class of multivariate GARCH models suitable for large $N$ are factor models.
They are characterized by using constant $P_k$ with rank 1 obtained from $C$ by equation
\myref{a6} with the $K$ largest eigenvalues. A model structure similar to O-GARCH \citep{Alex} could be achieved 
by setting
$A_k=\sqrt{\alpha_k}P_k$ and $G_k=\sqrt{\gamma_k}P_k$ with time-dependent 
$\lambda_k(t)$ in $H(t)=\sum \lambda_k(t) P_k$. 
 A slightly more complex approach are full-factor models like \cite{vront} where the conditional correlations obey a GARCH behavior, a feature which has some similarity with the behavior of our dynamic $\beta$.

Another  model is the GO-GARCH \citep{GOGarch}, which
is applied to the de-correlated returns $\tilde{r}(t)=C^{-1/2} \; r(t)$. The $K$ 
principal components of $\tilde{r}$ are rotated by a $K-$dimensional rotation.  The additional
$K(K-1)/2$ angles are additional parameters in MLE.  Our model therefore may be called an effective two-factor model.
It differs from these factor models by two features. The first is that we have always two factors. The factor $v_0(t)$ describes the market and we can interpret the nonmarket
component $v_1(t)$ as a second factor that corresponds to $N-1$ identical factors $v_1(t)$.   Second, we use a time-dependent
$P_k(t)$. Equation \myref{a13} (or equation \myref{R2} in appendix \ref{recurs}) 
corresponds to a time-dependent 
rotation with angle $\cos(\phi(t))=m(t)$ as in a $K=2$ GO-GARCH. However, 
$\phi(t)$ is  determined by the recursion and not by estimation. 

For the large literature on other factor ARCH, GARCH and DCC models the reader is referred to \cite{bolarch} and \cite{DCCL}.

Another class of models avoids numerical calculation of $H^{-1}(t)$ by a restricted $H(t)$ as 
in our model. An example is the DECO model of \cite{deco}. In the conditional
correlation matrix $R(t)$  pairwise correlations are equal
\begin{equation} \mylabel{a16}
R(t)=(1-\rho(t))\; I\; +\;\rho(t)
\end{equation}
which allows analytical calculation of $R^{-1}(t)$. In this model first the diagonal 
matrix of conditional expectation values $D_i^2(t)$ of $r_i^2(t)$ is
determined by $N$ univariate GARCH models. With this one can then calculate \mbox{$H(t)=D(t) R(t)  D(t)$} and obtain the average correlation $\rho(t)$ from the DCC recursion.
Our model could reproduce a DECO type model by setting $\beta (t)=1$  with
\begin{equation} \mylabel{a17}
\rho(t)=\left(1+\frac{v_1(t)}{v_0(t)}\right)^{-1}
\end{equation}

\section{Estimation and results for S\&P stocks \label{fits}}

\begin{figure}[htb]
\begin{minipage}[t]{0.5\linewidth}
\includegraphics[width=\linewidth]{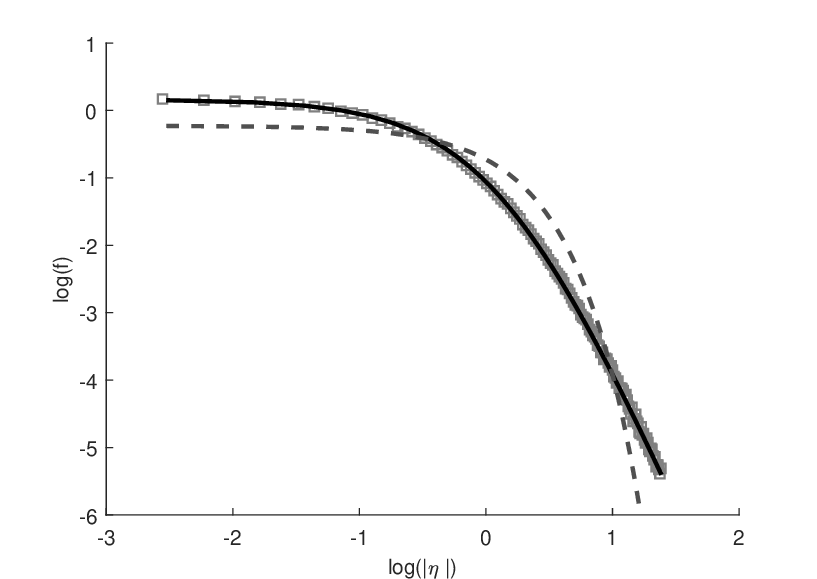}
\end{minipage}
\begin{minipage}[t]{0.5\linewidth}
\includegraphics[width=\linewidth]{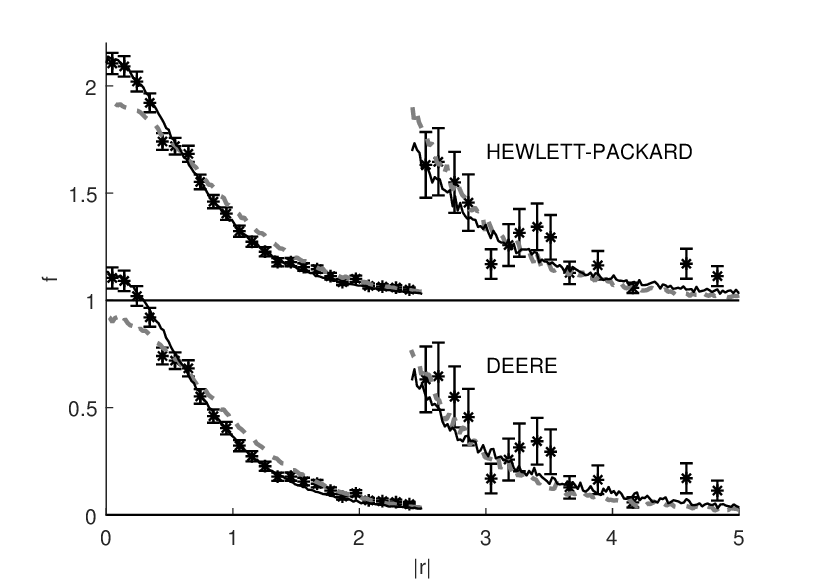}
\end{minipage}
\caption{\label{degarch}\em The left panel shows the pdf of all GARCH filtered returns
          $\eta_G(t)_i$ obtained from the raw returns by applying \myref{a12a}.
        The black line corresponds to a fitted t-distribution with $\nu=3.32$,
        the broken gray line shows a Gaussian distribution.
          In the right panel two typical pdfs for stocks of the S\&P market
          are compared with the predicted pdf using t-distributed (black) or
          Gaussian (gray) noise (see appendix \ref{sec:chi} for calculation of errors). 
          The values (log of the pdf) $|r| > 2.5$
          are multiplied by a factor of 10. The theoretical curves are based
          on the 2-parameter fits as shown in table \ref{tab_fit}.}
\end{figure}

\subsection{Preliminary fit and noise dependence}

In this section we describe the maximum likelihood estimation with data of
daily returns of 356 stocks that were constituents of the S\&P index for the years 1995-2013.
For our analysis we use data from Thompson Reuters on the closing price of stocks which were 
continuously traded with sufficient volume
throughout the sample period and had a meaningful market capitalization.\footnote{We excluded stocks which price did not change for more then 8 \% of the trading days, or which were exempt from trading or for which no trading was recorded for more than 10 days in a row.  We manually deleted 15 stocks which price movements at some point showed similarities to penny stocks and/or which market capitalization  was very low.}

The scale of returns is arbitrary. We choose to normalize $r(t)$ such that the average of $r_i^2(t)$ over all $i$ and $t$ is equal to one. This affects the scale in figures \ref{degarch}, \ref{fig5rm}, \ref{pdf_chisq}, \ref{fig:cov} and \ref{fig:5dist}.

\begin{table}[htb]
\begin{center}
\begin{tabular}{c|c |c c c c c c} \hline
\hline
$N_{par}$ & $ L/T$ & $\alpha_{0}\cdot 10^{2}$ & $\gamma_{0}\cdot 10^{2}$ & $\alpha_{1}\cdot 10^{1} $
       & $\gamma_{1}\cdot 10^{2}$ & $\alpha_{01}\cdot 10^{2}$ & $\gamma_{01}\cdot 10^{2}$
       \\
        \hline

2 & -52.2 & 4.871  & 0.383  & $\alpha_{0}$ & $\gamma_{0}$ & $\alpha_{0}$ & $\gamma_{0}$\\
 &       &  \small{(0.068)}  & \small{(0.014)}  &               &               &               &   \\
2 & -2.57 & 3.132  & 0.284  & $\alpha_{0}$ & $\gamma_{0}$ & $\alpha_{0}$ & $\gamma_{0}$\\
  &       & \small{(0.048)}  & \small{(0.011)}  &               &               &               & \\
4 & -0.10 & 1.592  & 0.328  & 2.472 & 0.766 & $\alpha_{0}$ & $\gamma_{0}$ \\
  &       & \small{(0.042)}  & \small{(0.016)}  & \small{(0.050)} & \small{(0.078)} &               &                \\
6 & 0.00  & 5.14   & 4.13   & 2.487 & 0.781 & 1.673 & 0.298 \\
  &       & \small{(0.16)}   & \small{(0.31)}   & \small{(0.041)} & \small{(0.065)} & \small{(0.044)} & \small{(0.014)} \\ \hline
\end{tabular}
\caption{\em Maximum likelihood estimates of the parameters for the S\&P market.
         Column one states the number of parameters, column two the values of the
         log likelihood per time relative to the six parameter fit.
         Standard  errors are given in parentheses.
         The results in the two top rows are obtained with Gaussian noise,
         in rows 3-8 with t-distributed noise.}
\label{tab_fit}
\end{center}
\end{table}

We calculate the log likelihood $L$  of our model with the exact solution of the
recursion as described in appendix \ref{recurs}.
The initial  values $\bar{v}_\nu$ and $\bar{\beta}$ have been determined from
the observed covariance matrix $C$ in the first 4 years.
In a preliminary fit we  use only two parameters with $\alpha_{1}=\alpha_{0}$,
$\alpha_{01}=\alpha_{0}$, $\gamma_{1}=\gamma_{0}$ and
$\gamma_{01}=\gamma_{0}$ together with a Gaussian distributed noise.
The resulting log likelihood (divided by $T$) and the parameter values
are given in table \ref{tab_fit}.

As a check on the property of this fit we consider the so-called de-garched returns
$\eta_G(t)$. These are obtained by inverting equation \myref{a0}

\begin{equation} \mylabel{a12a}
\eta_G(t)=H(t)^{-1/2} \; r(t)
\end{equation}

If the \G model represents the data exactly $\eta_G(t)$ should be again Gaussian
distributed. This is not the case, as the pdf in the left panel of 
 figure \ref{degarch} shows. The pdf for $\eta_G(t)$ is well described by a
Student's t-distribution with a tail index of $\nu=3.32$.
This motivates to repeat the estimate with a t-distributed noise. The estimated
value of $\nu=3.25$ agrees within the errors with the value obtained before (see figure
\ref{degarch}). The resulting $L$ ($2^{nd}$ column in table \ref{tab_fit}) corresponds
to an astronomical increase of probability. Even the pessimistic evaluation using
the probability change per $t$ given by $exp(-\Delta L/T)$ is highly significant.

Such  a strong noise dependence is not present in applications of 
              univariate GARCH(1,1) models. There, the tail index can be reproduced either by 
              t-distributed or Gaussian noise. However,  since our model (hereafter called RMG for 
              restricted  matrix GARCH) has only four GARCH parameters, $\alpha_{0,1}$ and
              $\gamma_{0,1}$, the tail cannot be matched for all $N$ stocks when Gaussian noise is used \citep[see also][for a non-parametric moment analysis of stocks and indices]{WMA}.
              
              In the right panel of figure
\ref{degarch} two typical pdfs of returns are compared with the predicted density
either using t-distributed (black line) or Gaussian (broken gray line) noise. The latter
describes the tails but fails grossly for the Gaussian region near $r=0$ and in
the transition to the tail. Therefore, in all subsequent estimations we use a t-distributed
noise with  $\nu=3.35$.
              
\subsection{Using simulated returns to evaluate the fit}  

\begin{figure}[p]
    \begin{minipage}[t]{0.48\linewidth}
        \begin{center}
        \includegraphics[width=\linewidth, trim= 20 20 10 10, clip=true]{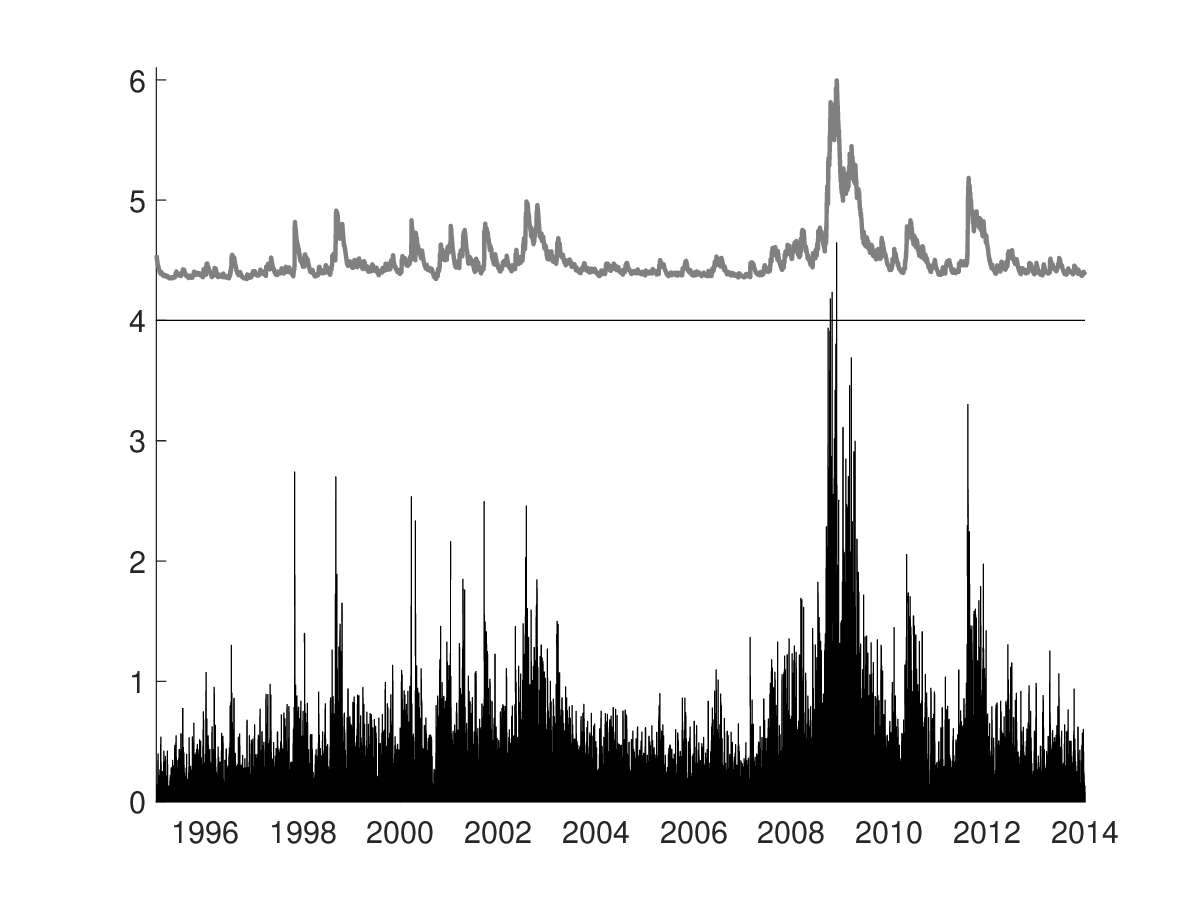}
        \end{center}
        \caption{\label{fig5rm} \em Market factor $\sqrt{v_0(t)}$ (top) and
           market return $|r_{M(t)}|$ (bottom). A value of 2 is added to $\sqrt{v_0(t)}$}
     \end{minipage}
\hspace*{0.02\linewidth}
    \begin{minipage}[t]{0.48\linewidth}
        \begin{center}
        \includegraphics[width=\linewidth, trim= 20 20 20 20, clip=true]{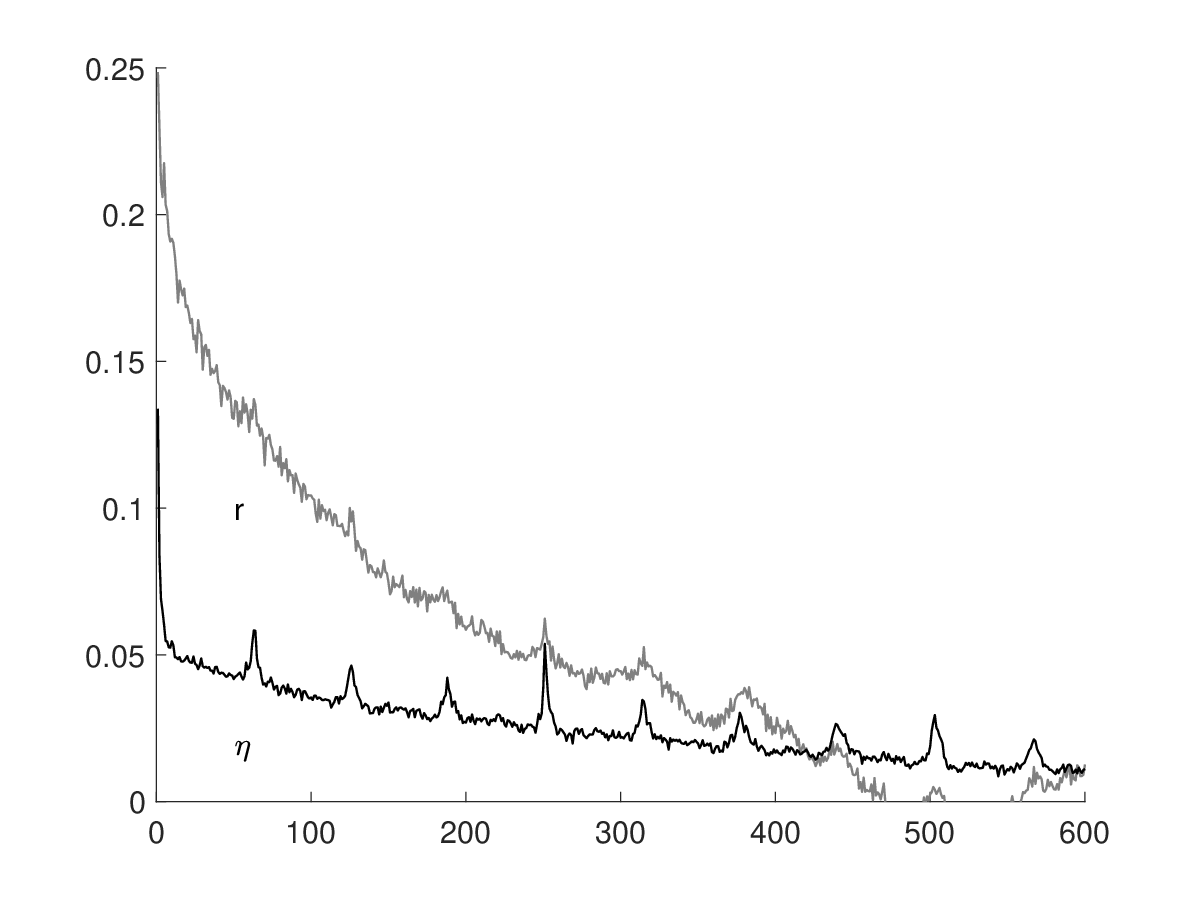}
        \end{center}
        \caption{\label{fig6degarch} \em Autocorrelation for returns $|r_{i}(t)|$
           averaged over all stocks (top)
           and the averaged GARCH filtered returns $|\eta_G(ti)|$ (bottom).}
        \end{minipage}
\end{figure}

\begin{figure}[p]
\begin{center}
\includegraphics[width=0.93\linewidth, trim= 50 35 50 15, clip=true]{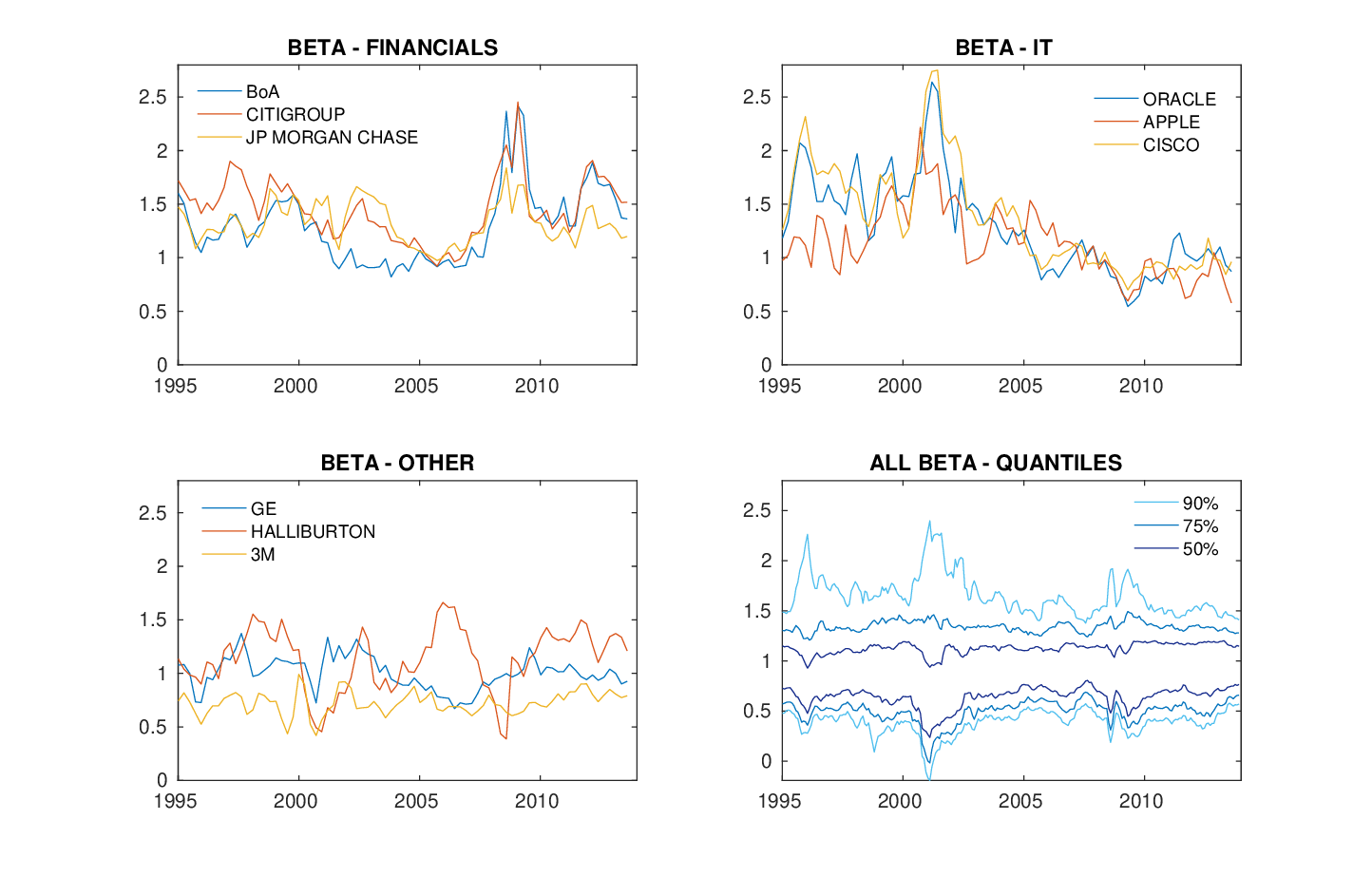}
\end{center}
\caption{\label{figbeta} \em Beta values from the RMG model. The top panels and the bottom left panel show beta values for different stocks, averaged over a 60-day window. The bottom right panel shows the dynamics of the distribution of beta values by the quantiles, using a 20-day window.}
\end{figure}

To make sure that our model describes the distribution of returns adequately we chose to simulate the returns and to compare them to the empirical ones.            

 Equation \myref{a0} predicts $r(t)$ for given $H(t)$ from any GARCH model and noise at time $t$. This relationship can easily be utilized for a Monte Carlo simulation. By multiplying $H(t)^{-1/2}$ with a vector of noise we obtain simulated predicted returns $\hat{r}(t)$.
When we repeat this process with different noise $n_{sim}$ times we obtain a
range of possible values for the returns. By performing this procedure at all $t$ we obtain
$n_{sim} T$ predictions for $\hat{r}$. Individual time series cannot be obtained, however,
average values as pdf should agree with the observed quantities.

As we will see in section \ref{data}, this procedure can also be used to compare distributions of covariances for different models. To obtain smooth distributions we found that $n_{sim}=10$ is sufficient for the pdf of $\hat{r}$, covariances require $n_{sim}=40$).

The drawback of the preliminary two parameter fit consists in the small value for $\gamma_{0}$.
The estimate corresponds to autocorrelation that lasts for 339 days, which is roughly twice as long as we would expect from the observed individual $|r_{i}(t)|$. The likelihood improves when we add $\alpha_{01}=\alpha_{0}$ and $\gamma_{01}=\gamma_{0}$ to the estimation (third row in table \ref{tab_fit}).
However, only after including all six $\alpha$ and $\gamma$ parameters the likelihood increases again and
we obtain a reasonable value for $\gamma_{0}$. The estimates for the parameters that govern the dynamics of $\beta$ are smaller by comparison, indicating that $\beta(t)$ varies much less than the factors
$\sqrt{v_\nu(t)}$. As in any GARCH model, these exhibit much less fluctuations
than the underlying returns. This is shown in figure \ref{fig5rm} where the
market factor $\sqrt{v_0(t)}$ is compared with $r_{M}(t)$.

More details on the autocorrelation are shown in figure
 \ref{fig6degarch}, where we compare the
GARCH filtered returns $|\eta_G(t)_i|$ with $|r_{i}(t)|$ averaged over all
stocks $i$. 
The large statistics allow to resolve the peaks at multiples of three
months, which are caused by the dividend pay days. These correlations are outside of the
realm of any \G and survive the filtering. One can however see that the autocorrelation in $\eta$ is not perfectly removed. The reason is that with just 6 parameters the model gets the volatility right on average but mutes (inflates) stocks with very high (low) volatility since the $\beta$ move slower than $v_0$. As we will see in section \ref{sec:cov}, this particular weakness however does not carry over to the estimated covariances.


\subsection{Beta values for financials and IT companies}


Figure \ref{figbeta} finally gives an overview about the beta values that can be
obtained from the RMG. Note that these betas are normalized to $\beta'(t)\cdot \beta(t)=N$.
The top left panel shows the development of beta values for some financial stocks,
the top right shows the beta of IT-related companies. Stocks from both groups
show strong similarity within their group. The bottom left panel shows the beta
for stocks from other sectors. They develop much more diverse. The bottom
right panel illustrates how the distribution of beta values has develops over
time. At the times of the IT bubble and the Lehmann crisis we observe peaks for the
beta values but also a much wider distribution in general.



\section{Comparison with other GARCH models\label{data}} 
In the following we will compare the estimation results of our model with those of other
multivariate GARCH models. We will look at the predicted returns and covariances. Here we are interested in the practical implications of the dynamics of the model for  application. In particular we are interested in a rough approximation of the precision of the estimated covariances with respect to the empirical data. A formal model comparison and selection process would be outside the scope of this paper.
 
\subsection{Comparison of distributions of returns}

\begin{figure}[p]
\begin{minipage}[t]{0.5\linewidth}
\includegraphics[width=\linewidth]{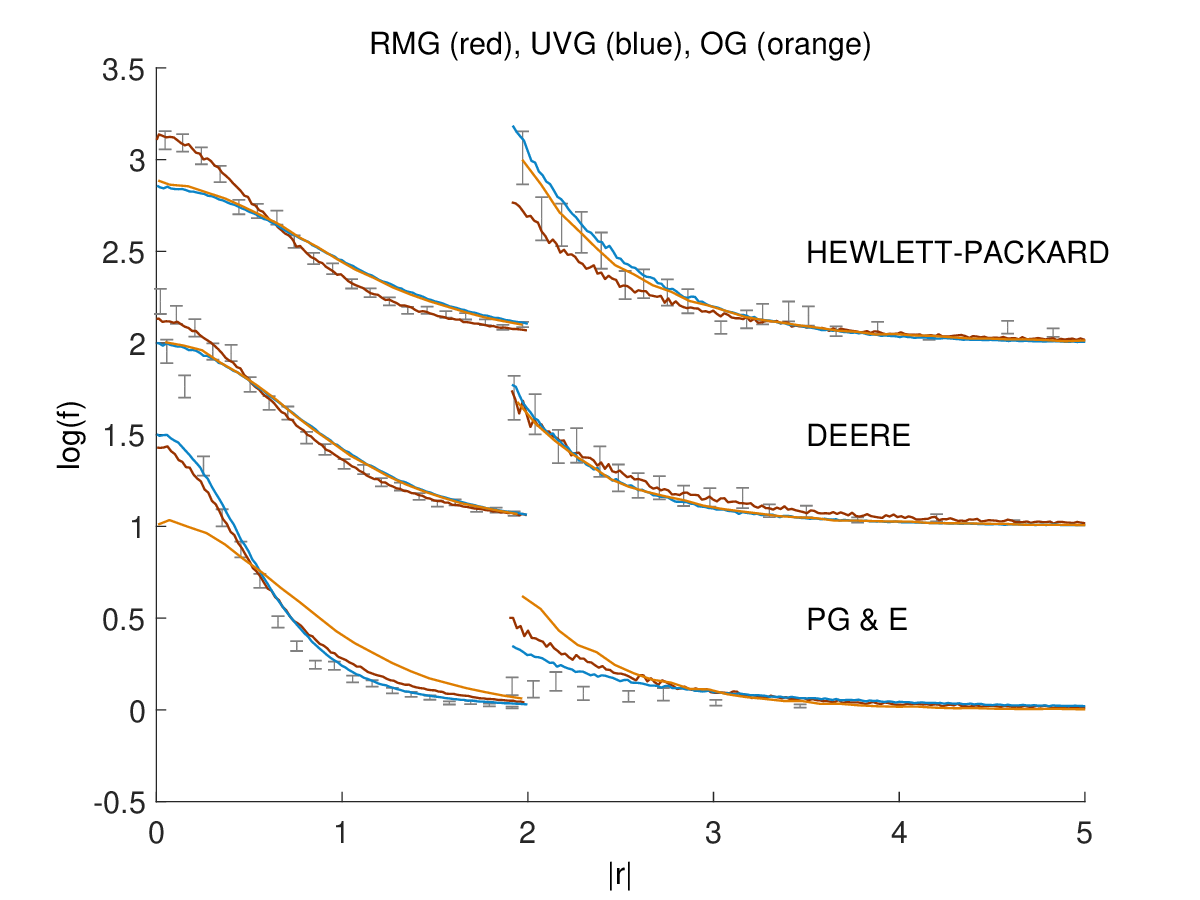}
\end{minipage}
\begin{minipage}[t]{0.5\linewidth}
\includegraphics[width=\linewidth]{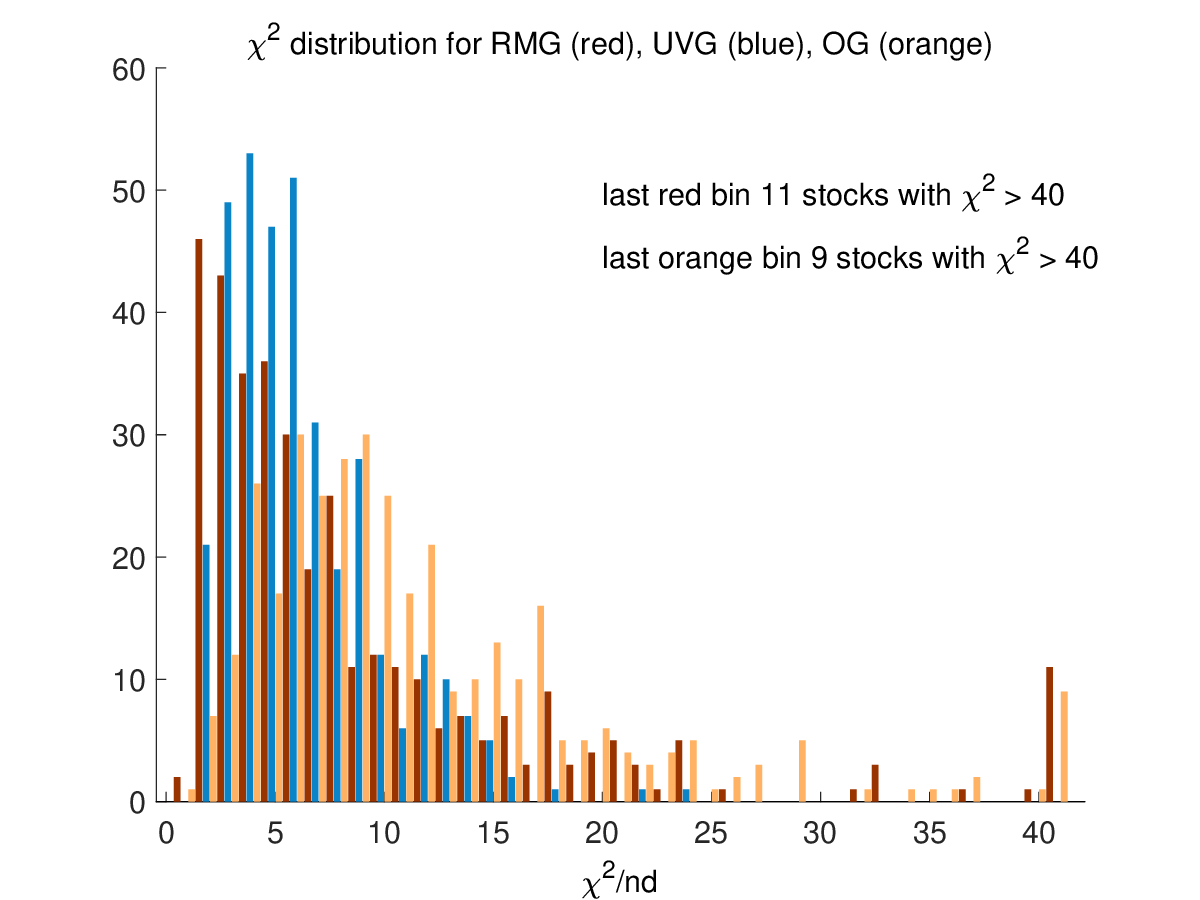}
\end{minipage}
\caption{\label{pdf_chisq}\em The left panel shows the pdf of returns compared with the
         prediction of RMG (red), UVG (blue) and OG (orange). The right panel
         shows a histogram of $\chi^2$  for all stocks for RMG, UVG and OG.}
\end{figure}

We start by comparing the predicted pdf of stock returns 
from the RMG model discussed in section \ref{fits} with those of
the univariate GARCH(1,1) (abbreviated UVG) and OGARCH model (hereafter abbreviated  OG). 
In both cases we use i.i.d. Gaussian $\varepsilon$ for the noise.

The left part of figure \ref{pdf_chisq} shows that the pdf for the same two stocks as in 
figure \ref{degarch} is in reasonable agreement with the three models. 
RMG and OG
fail in cases where the leading eigenvector does not dominate $H(t)$, as the 
third example for PG\&E shows.

For a more comprehensive comparison we use the $\chi^2/n_d$ ratio, which is the sum
of quadratic deviation in units of the squared error divided by the number $n_d$
of bins. A histogram that shows how well the distributions of predicted returns fit the data measured by these ratios for all 356 stocks of the S\&P market
is shown for RMG, OG and UVG in the right panel of figure \ref{pdf_chisq}.
All three distributions exhibit a peak around a value of 4-5 which
corresponds to the 5\% confidence level. Values between 10-20 lead still to a
qualitative description (see appendix \ref{sec:chi} for details on the calculation of the statistics). In contrast to UVG, both RMG and OG have a small fraction (5\%)
of outliers mainly from the energy sector as PG\&E. On average RMG performs
better than OG which may be due to the systematic error by covariance targeting.
We stress that for UVG and OG 712 parameter have to be determined. The only
six parameters used in RMG lead to a much more parsimonious description of the
data.

\subsection{Comparison of covariances}\label{sec:cov}

\begin{figure}[p]
\begin{center}
\includegraphics[width=\linewidth,trim= 40 400 30 35, clip=true]{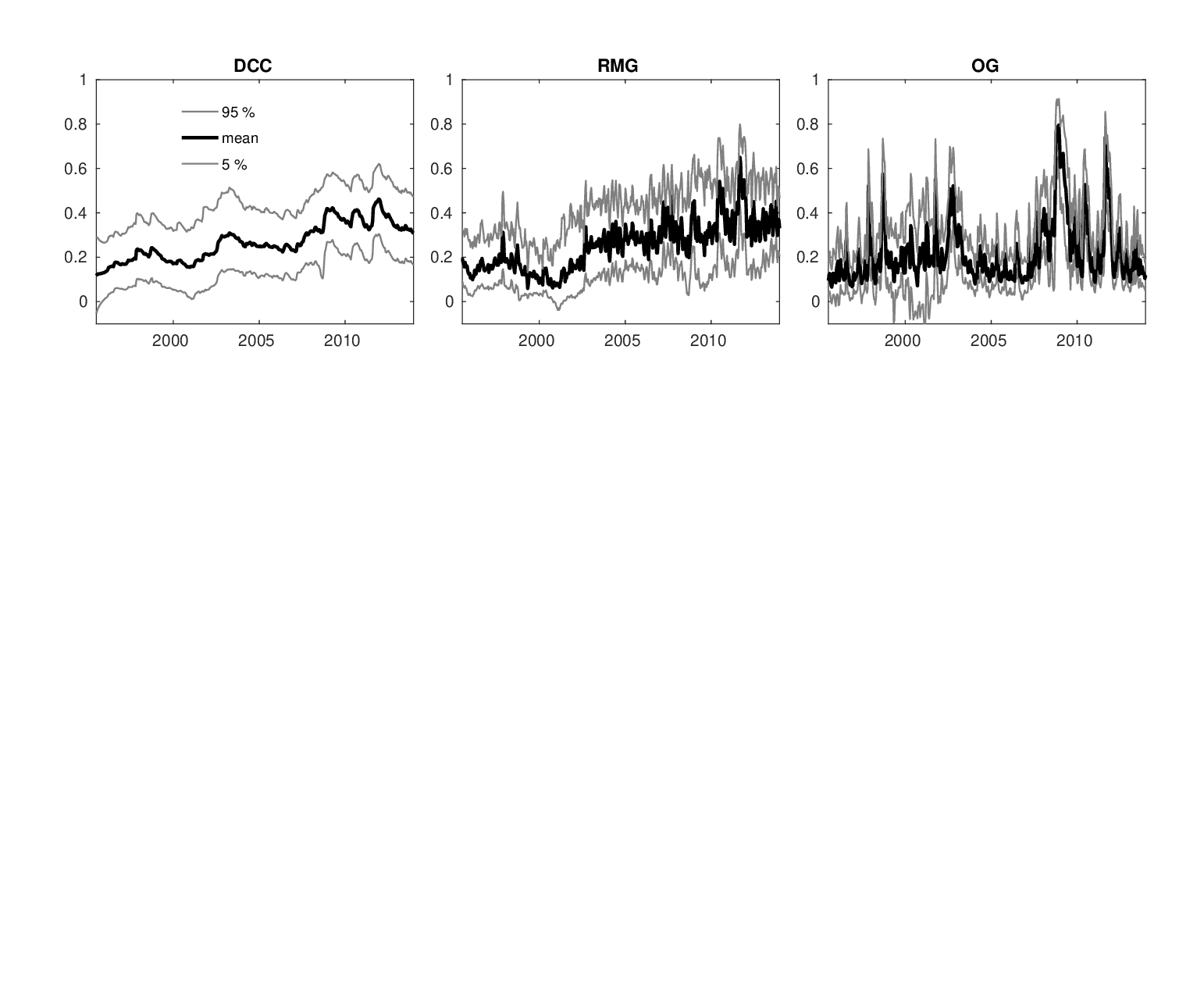}
\end{center}
\caption{\em Averaged  correlation as function of time. The figure shows the mean as well 
        as the 90\% bands of correlation coefficients obtained for the DCC, RMG, and 
        OG model. For the calculation of all correlations  10-day non-overlapping windows 
        have been applied.}\label{fig:4c}
\end{figure}

\begin{figure}[p]
\begin{center}
\includegraphics[width=\linewidth,trim= 15 220 20 20 , clip=true]{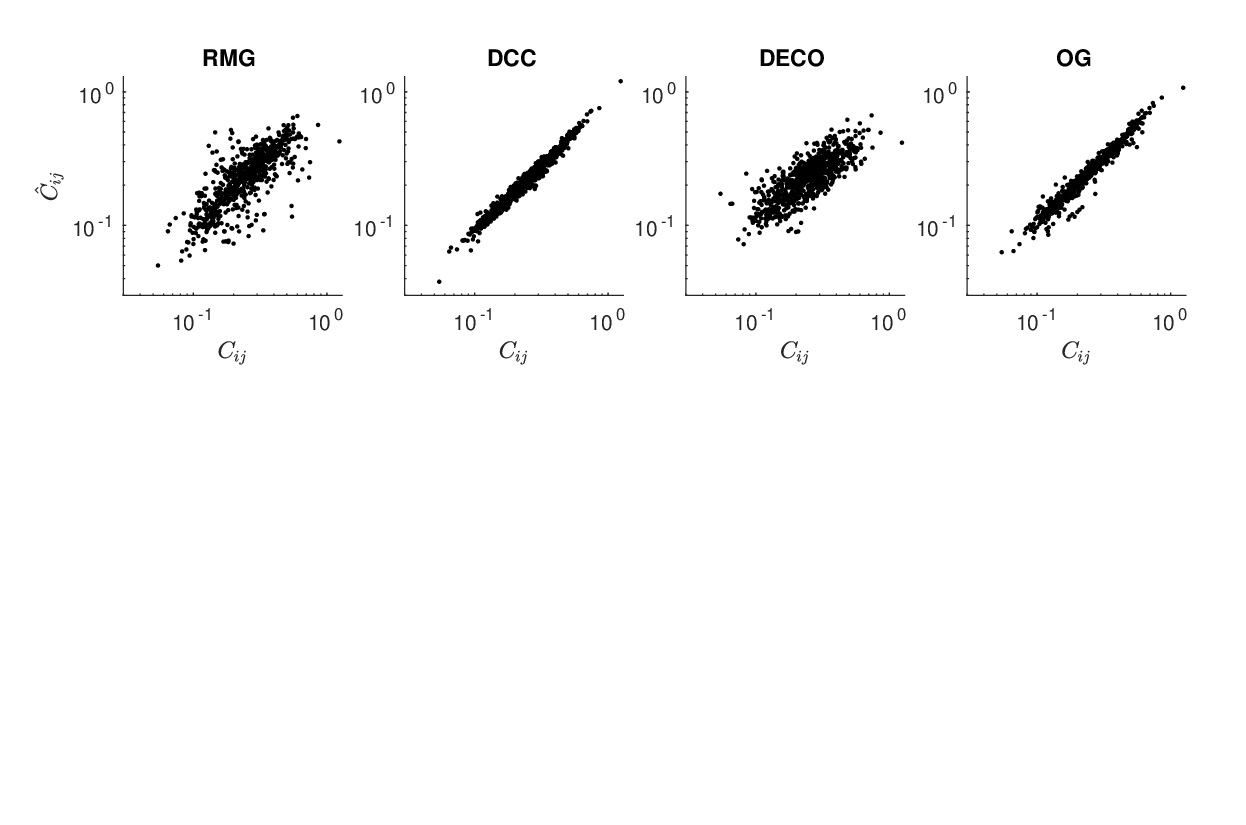}
\end{center}
\caption{\em Covariances of simulated returns versus covariances of empirical returns. For better visibility we plot $C_{ij}$ for only 792 randomly chosen pairs of stocks on a log-log scale.}
\label{fig:cov}
\end{figure}

In the following we compare the covariances obtained from the RMG model with those obtained from the DCC, OG and DECO model and those implied by the empirical data. 
The form of $H_t$ in RMG, OG and DECO is in each case in some way restricted, which influences the estimated covariances.
We therefore consider the standard DCC model, applied to subsamples of the data, as a benchmark estimation.

 Even if the DCC model allows relatively 
large $N$, it is not designed for a sample like ours. It is typically 
applied for (and works best) for $N \leq 10$ \citep[see][]{ESS,DCCL}. For approaches to overcome this problem by using modified estimators, composite likelihood and shrinkage methods we refer the reader to \cite{aielli} and \cite{large_dcc2}.

For computational reasons the comparisons in this section (with exception of figure \ref{fig:4c}) are based on subsamples of covariances.
We apply the  DCC model to blocks of 8 stocks. The division of the stocks into these blocks is random. With a total number of 44 blocks we cover 352 stocks and 1,232 covariances. 
The RMG, OG and DECO model have been estimated for the entire sample of stocks, the comparison of covariances however is always based on the exact same 1,232 covariances.
By repeating the sampling we have verified that the results presented in this section do not depend on the sampling.

Figure \ref{fig:4c} shows the average correlation $\rho(t)$ together with the 90\% 
interval for DCC, RMG and OG. The average correlation from DECO (not shown) is almost identical 
to that of the DCC model. The overall dynamics of the DCC and the RMG model seem to be 
rather similar. RMG exhibits larger fluctuations since it describes all pairs
with 6 parameters.
The OG model leads to  qualitatively very different dynamics of the correlations.

\begin{table}[htb]
\begin{center}
\begin{tabular}{c  |c c c c } \hline
\hline
   &  RMG & DECO & DCC & OG \\
        \hline
\\
   $ RMSE(\hat{C}_{ij}-C_{ij}) $    &
   2.1390 & 3.0633 & 1.4018 & 0.9513 \\
 $ <|\hat{C}_{ij}-C_{ij})| >$    &
 6.88 $\%$ & 11.03 $\%$ & 5.97 $\%$ & 4.44 $\%$ \\
 \\
 \hline
 & & &  & \\
 $<\Delta (\hat{r}_i \hat{r}_j,r_ir_j)>$  & 0.0091 & 0.0570 & 0.0558 & 0.1395  \\
\small $ std(\cdot)$ &
 \small 0.079 & \small 0.031 & \small 0.031 & \small 0.054 \\
 $ <|\Delta (\hat{r}_i \hat{r}_j,r_ir_j)|>$   & 0.0650 & 0.0575 & 0.0562 & 0.1395 \\
\small $ std(\cdot) $ &
 \small 0.046 & \small 0.030 & \small 0.030  & \small 0.054 \\
 \\

\hline
\end{tabular}
\caption{\em The upper part of the table shows the differences between the estimated versus the empirical covariances. We also report the average percentage deviation of this difference with respect to $C_{ij}$.
The bottom part of the table reports the differences in the distributions of the covariances, applying Cliff's delta test. The first line gives the average value, the second line the average absolute difference together with the standard deviation.}
        \label{tab_dtest}
\end{center}
\end{table}

We can judge whether the models lead to a correct description of the covariances  by confronting
them with the empirical data. For the conditional covariances perform
the Monte Carlo simulation described in section \ref{fits} to obtain time averaged
covariances $\hat{C}_{ij}$ and distributions of $\hat{r}_i\hat{r}_j$. 

In figure \ref{fig:cov} 
we show scatterplots comparing $\hat{C}_{ij}$ for the models with the observed time
averaged covariances $C_{ij}$. Since $C$ is input in OG and is used for the mean
of $H$ in DCC by covariance targeting we observe almost straight lines broadened by noise for these two cases.
The somewhat larger deviation seen for RMG is partly due to the systematic error
of the random matrix $C$. In RMG only the properties of the large eigenvalue of $C$
are used. The performance of DECO is similar and is caused due by the assumed stock-independent correlation.
We can quantify this observation by calculating the differences between simulated and empirical covariances as $RMSE(C_{ij}-\hat{C}_{ij})$ given
in the top part of table \ref{tab_dtest}. 


A more interesting comparison is to look at the distributions of the covariances. These distributions have pronounced peaks around 0 which are very dominant but not very decisive for our comparison. We have therefore decided to look at absolute covariances, in particular $\log(|r_i r_j|)$ for \mbox{$|r_i r_j| > 0$}. Three examples for this distribution are shown in the figure in appendix \ref{sec:appcov}.

We measure the difference between the predicted distributions and those given by the data by calculating \cite{cliff}'s delta,  denoted by $\Delta$.
This statistic relies on comparing the elements in both 
distributions and counting how often each element is larger (smaller) than any 
element from the other distribution. Delta is bound within [-1,1], a value of 0 
 signals distributions that are indistinguishable while positive (negative) 
values signal some degree of imperfection in the overlap of the two distributions.

\begin{equation}
\Delta_{ij} = \frac{1}{n_t n_\tau} \sum_{t,\tau}  \boldsymbol{[} |\hat{r}_i \hat{r}_j(\tau)| > |r_i r_j(t)| \boldsymbol{]}
- \boldsymbol{[}|\hat{r}_i \hat{r}_j(\tau)| < |r_i r_j(t)| \boldsymbol{]}
\end{equation}

In the bottom part of table \ref{tab_dtest} we report the mean of $\Delta$ and 
$|\Delta|$. Their standard deviation are a measure of the uncertainty. RMG has a $\Delta$ 
compatible with zero, while the values of DECO and DCC are slightly larger than zero. The dispersion of $\Delta$ however is larger for RMG than for the other models. Judging by the absolute $\Delta$ we can see that RMG achieves results that are only marginally behind DECO and DCC. The OG model for comparison performs noticeably worse. 

\section{Applications \label{Applications}} 

\subsection{Market transition}
The large number of stocks in our sample allows to search for group specific regularities. 
An obvious question is if the correlations that can be extracted from  
$H(t)$ differ for stocks from
specific sectors, and more interestingly, how they develop over time. 
Our sample period covers a time period during which we
have seen pronounced changes in the market, including the IT bubble and the 
financial crisis. The visual inspection of some stocks' betas in 
section \ref{data} has already hinted at changes in the IT and the financial 
sector, which we want to analyze in more detail.
For this reason we use the estimated $H(t)$  and then derive the conditional
correlation matrices from $D^{-1} H(t)  D^{-1}$ ,
where $D$ is a diagonal matrix with the square root of $H_{ii}(t)$ on the 
main diagonal. We can use these correlations to
calculate the median correlation for pairs of stocks from specific sectors (GICS classification).

\begin{figure}[htbp]
\begin{center}
\includegraphics[width=0.95\linewidth]{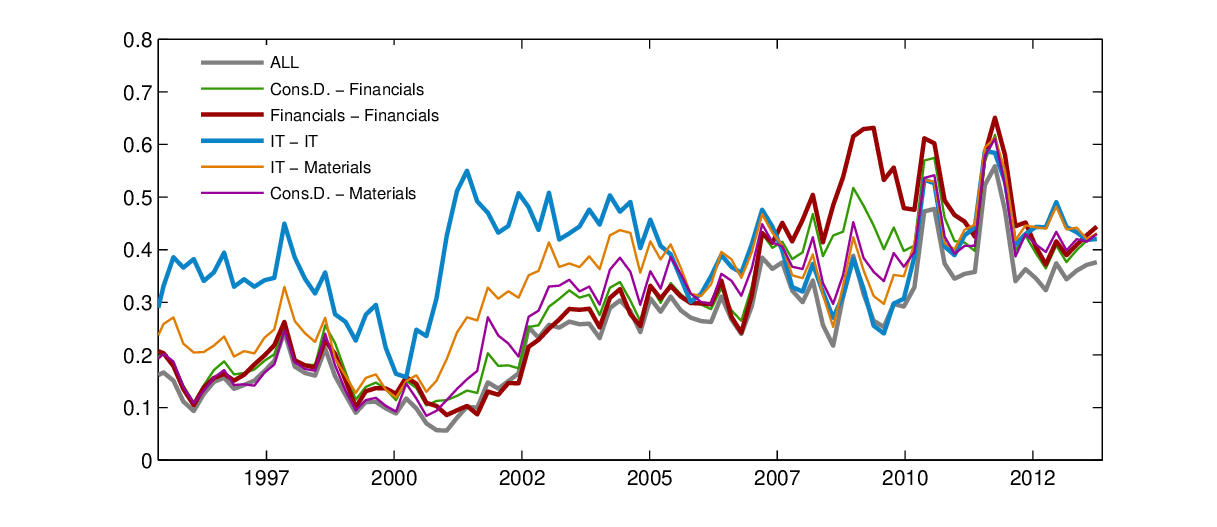}
\end{center}
\caption{\label{fig:Htdyn}\em Development of median correlation by sector, averaged over 50 days. We show the average of all stock correlations (bold gray line), as well as some of the sectors with the strongest correlations. The sector with the strongest inter-sector correlation has for a long time been the IT sector (blue), later the financial sector (red) has taken over this role.}
\end{figure}

Figure \ref{fig:Htdyn} shows some of these median correlations. 
We observe very high average within-sector-correlations for stocks in the IT and financial sector.
 But also 
some correlations between stocks of different sectors are rather high, for example when the consumer or materials sectors are involved.

In general we observe two important changes over time. First, the overall level of correlations has shifted upwards from 2002 until 2007. The
second observation is that the relative contribution of different sectors to overall correlation has changed. The most important change is that the financial sector surpasses the IT sector in terms of correlation
around 2006.

These changes can be analyzed in more detail.
In the analysis of \cite{Radd} of the US, the UK, and the German stock market the same
change has been found in the behavior based on the stock's beta values,
 derived under the assumption that
              the covariance matrix of the returns has one large eigenvalue already
              at a time window sizes of 3 years. In the years 1994--2006 stocks with high trading volume and high beta mainly came from the information technology sector, whereas in 2006--2012 such stocks came mainly from the 
              financial sector. Since a $\beta>1$ signals a risky investment,
a market risk measure $\hat{R}(t,s)$ has been defined for the sectors $s$ by multiplying
$\beta_i>1$ with the number $V(t,i)$ of traded shares in each time window.
\begin{equation} \mylabel{b14}
\hat{R}(t,s)=A_S \sum_{i \epsilon s}\theta(\beta_i-1.0)\beta_i(t)\; V(t,i)
\end{equation}

\begin{figure}[htbp]
\begin{center}
 \includegraphics[width=0.95\linewidth]{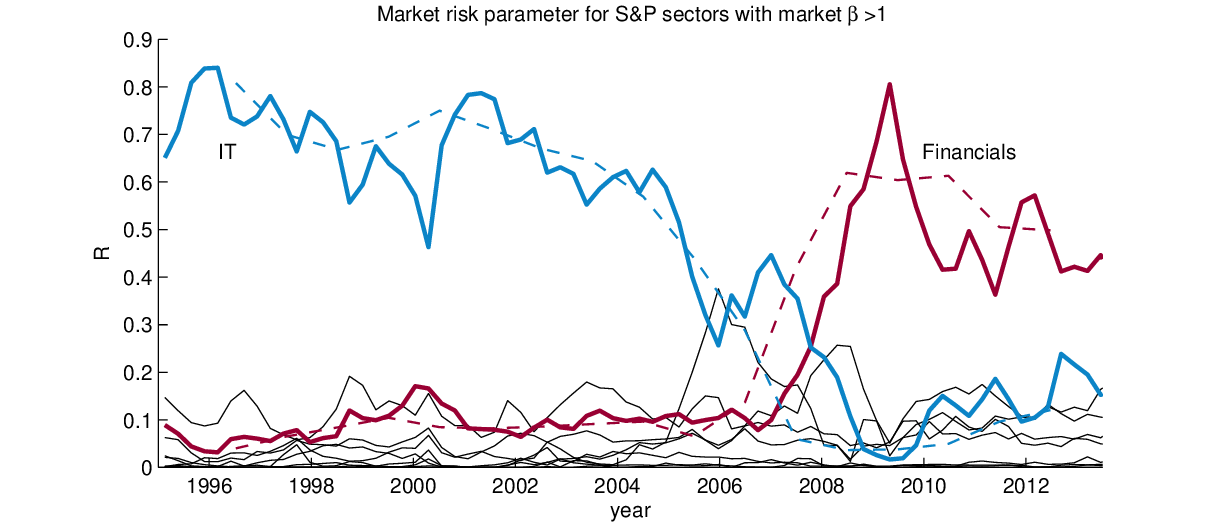}
	\end{center}
  \caption{\label{fig7risk} \em Time dependence of the risk parameter described in equation \myref{b14}
           for the sectors of the S\&P market. The dashed line shows the result of
           \cite{Radd}. The solid lines (blue for the information technology and red for the financial
           sector) use daily $\beta$s from the RMG.}
\end{figure}

The normalization constant $A_S$ is chosen to have $\sum_s\; \hat{R}(t,s)=1$.
When we apply this measure we see  that before 2006 only the information technology sector
and after 2006 only the financial sector exhibit large values of the risk 
measure. 

However, the time and the duration of this transition had a
systematic error of 1.5 years due to the window size. Repeating the calculation of $\hat{R}$ with
the $\beta$ obtained from the RMG model serves two purposes. Firstly it is a check whether 
in RMG the market property can be reproduced and secondly the transition
time can be determined more accurately, since daily $\beta$ are known from
RMG. To reduce the noise on $\beta$ we average $\hat{R}(t,s)$ over one month.
In figure \ref{fig7risk} the risk parameters from equation \myref{b14} for the
S\&P market is shown as a function of time. The agreement with the
previous determination (dashed lines)
is good. With a time resolution of one month we can now safely say that the transition happens
during the year 2006.

\subsection{Leverage effect}

The leverage effect consists in a negative correlation between 
volatility and future returns \citep{black2,lev2}. It is a relatively small
effect \citep{Schwert}, but important for the estimation of risk. Since 
GARCH models provide a measurement of the daily volatility they are
well suited for an analysis of this effect.

In a first step we
determine for each stock the time correlation $L_i(t)$ between the
market volatility $v_0$ and the observed returns $r_{ti}$:
\begin{equation} \mylabel{b15}
L_i(t)=\frac{1}{N_L}\sum_{t'} \left(v_0(t'-t)-\bar{v_0}\right)r_{t'i}
\end{equation}
with the normalization factor $N_L^2=T\cdot var(v_0)\sum_t r_{ti}^2$. 
Empirically the $L_i(t) \cdot sign(t)$ are dominantly negative with large fluctuations.
To improve the sensitivity we use the asymmetry 
defined by
\begin{equation} \mylabel{b16}
A_i=\frac{1}{t_{m}}\sum_{t=1}^{t_{m}}\big(L_i(t)-L_i(-t)\big)
\end{equation}
with a maximum of the time lag $t_{m}$ of two months. $A_i$ corresponds to the 
difference in the area under $L(t)$ for positive and negative $t$. Due to the
time symmetry of our model, using either $r_{ti}$ from equation \myref{a0} or the filtered
returns $\eta_G$ should eliminate the leverage effect.

\begin{figure}[htbp]
\begin{minipage}[t]{0.5 \linewidth} 
 \includegraphics[width=\linewidth]{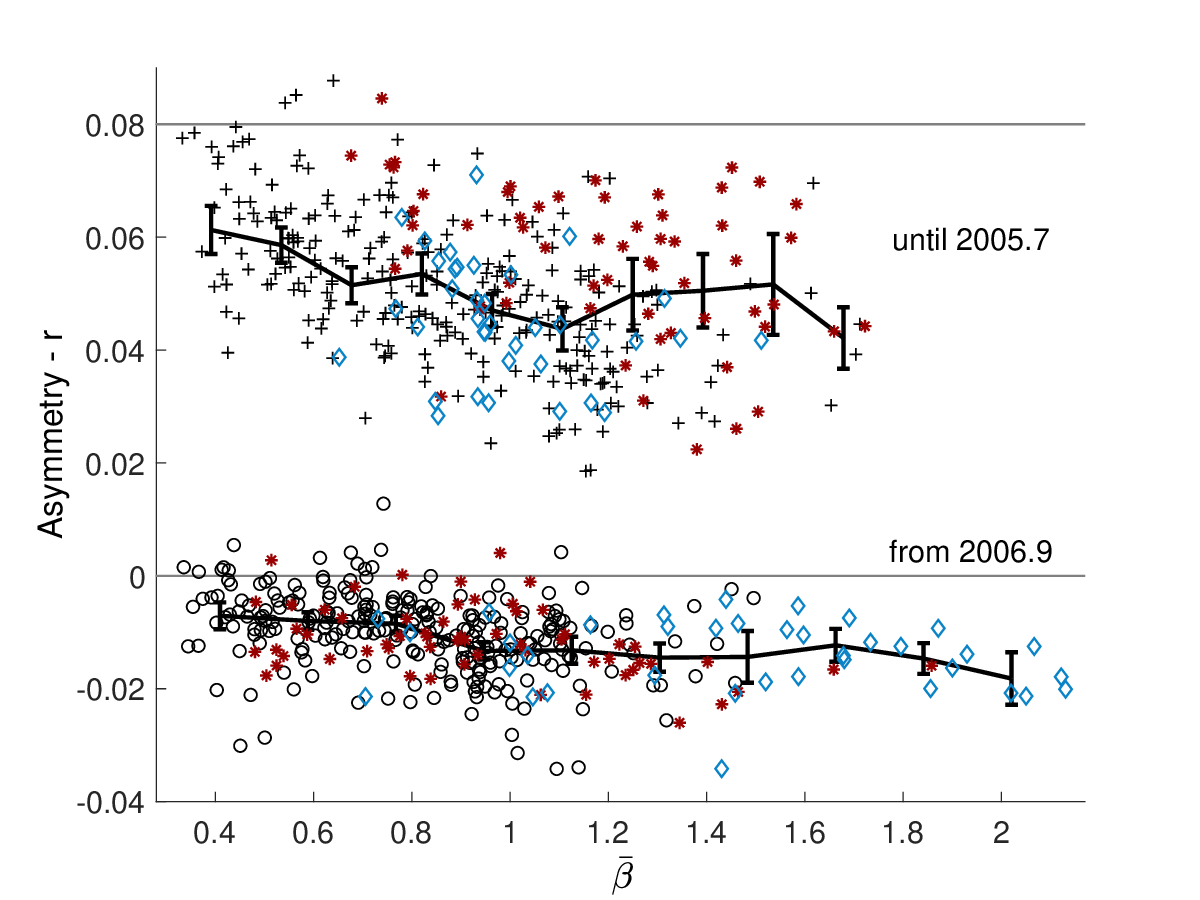}
\end{minipage}
\begin{minipage}[t]{0.5 \linewidth} 
 \includegraphics[width=\linewidth]{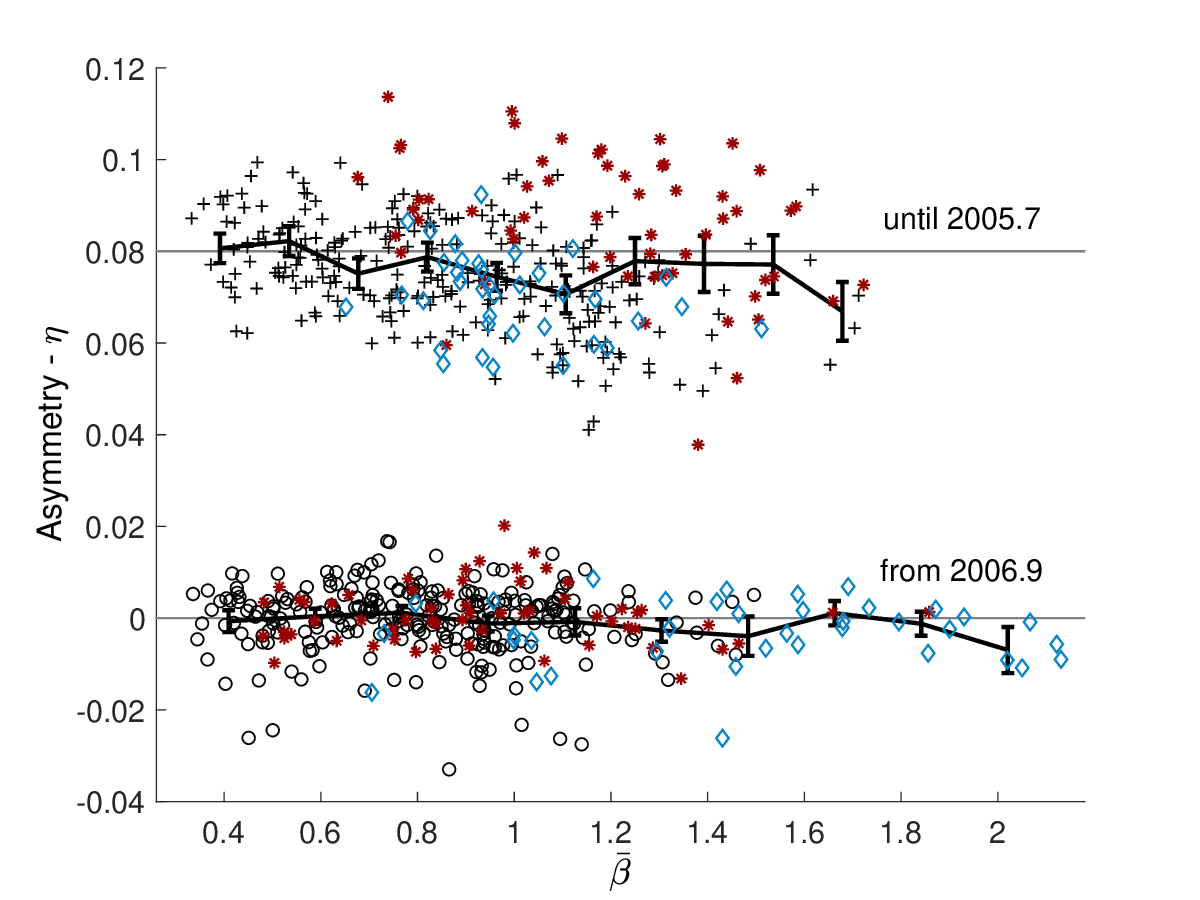}
 \end{minipage}
   \caption{\label{fig9}\em  Asymmetry of $r$ vs $\bar{\beta}$
    for stocks (left panel) and asymmetry of $\eta$ vs $\bar{\beta}$ (right panel)
     for all stocks. Blue diamonds refer to stocks from the technology sector, red stars to financials.
     The black line shows the average $\bar{\beta}$ and standard deviation. In both panels the upper part shows the result for the time period 1995--2005 and the lower part the results for 2007--2013. }
\end{figure}

It has been suggested by Black that the leverage effect is related to risk.
To test this suggestion we show in the left panel of figure \ref{fig9} the asymmetry as a
function of the mean value $\bar{\beta}_i$. Since ${\beta(t)}$ changes 
around 2006 we use two time series, one covering the years 1995 until 2005 and a second one
 covering 2007 until 2013. The asymmetries are clearly negative and increase
slightly with $\bar{\beta}$ as indicated by the line connecting the mean of $A$.
By replacing $r_{t}$ by $\eta_G$ in equation \myref{b15} the effect should disappear.
As shown in the right panel this is in fact the case.

The leverage effect can be included in the recursion. Analogous to
the GJR-GARCH model \citep{GJR} an  additional matrix proportional to
$P_\nu(t) \; GJR \; P_{\nu'}(t)$ with $GJR_{ij}= \delta_{ij}r_{ti}|r_{ti}|$ 
can be added on the r.h.s. of equation \myref{R1}.
For large $N$ the  recursions for $v_0$ and $\beta$ are unchanged, only
$v_1$ is affected. We repeated the fit including such a term. The likelihood
improves, however the values of $\alpha$ and $\gamma$ are inside the errors
the same. Also the results contained in figure \ref{fig9}
remain the same.

\section{Conclusions \label{concl}}

In our GARCH model we split the return into a market component and the remainder 
similar to a two-factor model. The parametrization of the covariance matrix
$H$ avoids the numerical inversion of $H$. The small number of parameters and
a computing time $T_{comp}\propto N$ allow the application to markets with a
large number of stocks. The off-diagonal elements of our $H(t)$ have a precision that is competitive with other established models.
Further, our model has the advantage that daily $\beta$ values 
relative to the market are determined.
We found that replacing
Gaussian by t-distributed noise is essential for the quality of the estimation results.

A possible development of the model would be to follow a similar strategy like DECO and to estimate univariate $h(t)$ in a first stage and to base the actual estimation of the model on residuals in a second stage.

\bibliographystyle{plainnat} 
\bibliography{mvg}
\section*{Appendix}
\begin{appendix}
\section{Derivation of the recursions   \label{recurs}}
We calculate $v_k(t+1)$ and $\beta(t+1)$ contained in the conditional covariance matrix
\begin{equation} \mylabel{A1}
H(t+1)=Nv_0(t+1)P_0(t+1)+v_1(t+1)(I-P_0(t+1))
\end{equation}
from the recursion \myref{R}
\begin{eqnarray} \mylabel{Ra}
H(t+1)&=& H(t)+  \sum_{k=0,1}P_{k}(t)\left[\alpha_{k} \left(r(t) r'(t)-H(t)\right)
   +\gamma_{k} \left(\bar{H}-H(t)\right)\right]P_{k}(t) \nonumber \\
  &+& P_{0}(t) \left[\alpha_{10} r(t) r'(t)+\gamma_{10}\bar{H}\right] P_{1}(t)
   \nonumber \\ &+&
  \;P_{1}(t) \left[\alpha_{10} r(t) r'(t)+\gamma_{10}\bar{H}\right] P_{0}(t)
\end{eqnarray}
The mean $\bar{H}$ is given by equation \myref{ab9}.
We apply the operation trace$P_{k}(t)\cdot$ on both sides of equation \myref{Ra}. The computation
can be done using the algebraic properties of $P_{k}$ shown in equations \myref{a3} and \myref{a4}. 
This leads to
\begin{eqnarray} \mylabel{A2}
\mbox{trace}H(t+1)P_{0}(t)&=& Nm^2(t)\; v_0(t+1)+{v_1(t+1)}(1-m^2(t))
    \nonumber \\ &=& N\; R_0(t)
\end{eqnarray}
\begin{eqnarray} \mylabel{A3}
\mbox{trace}H(t+1)P_1(t)&=& N\;(1-m^2(t))\; v_0(t+1)+{v_1(t+1)}(N-2+m^2(t))
     \nonumber \\ &=& N\; R_1(t)
\end{eqnarray}
where $m(t)=\beta'(t+1)\cdot \beta(t)/N$ denotes the overlap between $\beta(t)$ and
$\beta(t+1)$. The $R_k(t)$ represent the result of the same operation on 
the r.h.s. of equation \myref{Ra} involving only the diagonal terms given by
\begin{equation} \mylabel{A4}
R_0(t)=(1-\alpha_{0}-\gamma_{0})v_0(t)+\alpha_{0}r_M^2(t)+
     \frac{\gamma_{0}}{N}(N\bar{m}^2(t)\bar{v}_0-(1-\bar{m}^2(t))\bar{v}_1)
\end{equation}
\begin{eqnarray} \mylabel{A5}
R_1(t)&=&\frac{N-1}{N}(1-\alpha_{1}-\gamma_{1})v_1(t) + \alpha_{1}\bigl(\frac{r'(t) r(t)}{N}-r_M^2(t)\bigr)
     \nonumber \\ &+& 
    \frac{\gamma_{1}}{N}\bigl(N(1-\bar{m}^2(t))\bar{v}_0+(N-2+\bar{m}^2(t))\bar{v}_1 \bigr)
\end{eqnarray}
The determination of the mean values $\bar{v}_k$ and $\bar{\beta}$  from  $\bar{H}$ 
in equation \myref{ab9} is described in appendix \ref{noise}. $\bar{m}(t)$ denotes the overlap between
$\beta(t)$ and $\bar{\beta}$. Equations \myref{A3} and \myref{A4}  determine
$v_k(t+1)$ from the quantities $R_k(t)$ known at time $t$ and the overlap $m^2(t)$.
\begin{equation} \mylabel{R0}
v_0(t+1)=\frac{1}{Nm^2(t)-1}\; \left[(N-2+m^2(t))R_0-(1-m^2(t))R_1 \right]
\end{equation}
\begin{equation} \mylabel{R1}
v_1(t+1)=\frac{N}{Nm^2(t)-1}\; \left[m^2(t)R_1-(1-m^2(t))R_0 \right]
\end{equation}
To find $\beta(t+1)$ and thereby $m^2(t)$ we apply $H(t+1)$ to $\beta_i(t)$ and 
subtract \mbox{ $\beta_i(t)\; trace \; P_0(t)H(t+1)$ } to obtain 
\begin{eqnarray} \mylabel{A8}
\frac{1}{N}\bigl((H(t+1) \beta)_i - \beta_i (t)\; trace \; H(t+1) P_0 \bigr) &=& \nonumber \\
w(t+1)\bigl(\beta_i(t+1)-m(t)\beta_i(t)\bigr) &=& D_i(t)
\end{eqnarray}
where we used the abbreviation $w(t)=v_0(t)-v_1(t)/N$. $D_i(t)$ represents the same operation
on the r.h.s. of equation \myref{Ra} involving only the non-diagonal terms. It is given by
\begin{equation} \mylabel{A9}
D_i(t)=\alpha_{10}r_M(t)\bigl(r_i(t)-r_M(t)\beta_i(t)\bigr)+
    \gamma_{10}\;\bar{w}\;\bar{m}(t)\bigl(\bar{\beta}_i-\bar{m}(t)\beta_i(t)\bigr)
\end{equation}
This leads to the following recursion for $\beta(t+1)$.
\begin{equation} \mylabel{R2}
\beta_i(t+1)=m(t)\beta_i(t)+\frac{D_i(t)}{w(t+1)m(t)}
\end{equation}
The unknown $m(t)$ follows from the normalization of $\beta'(t+1) \cdot \beta(t+1)=N$.
By some miracle the equation for $m(t)$ is only quadratic in $m^2(t)$ despite the 
complicated structure of equations \myref{R0} and \myref{R1}. It reads
\begin{equation} \mylabel{Rb}
\left(R_0-\frac{R_0+R_1}{N}\right)^2 m^2(t)(1-m^2(t))=\frac{D'\cdot D}{N}\left(m^2(t)-\frac{1}{N}\right)^2
\end{equation}
One of the two solutions with $Nm^2(t)\le 1$ is to be excluded. Since $H(t+1)$ is insensitive to the
sign of $\beta(t+1)$ we always have a positive overlap $m(t)$. With the correct solution of $m(t)$
the set of equations \myref{R0},\myref{R1},\myref{R2} and \myref{Rb} yields $v_k(t+1)$ and $\beta(t+1)$
in terms of $r(t)$ and $\bar{H}$. The explicit solution looks rather complicated due to the 
dependence on $N$ in equation \myref{Rb}. For our purpose the limit $N \gg 1$ is sufficient. In this
case we have
\begin{equation} \mylabel{A10}
m^2(t) v_0(t+1)=R_0(t)
\end{equation}
With $D_i(t)$ from equation \myref{A9} we get $m(t)$ by
\begin{equation} \mylabel{A11}
m^2(t)=\left( 1+\frac{D'(t)\cdot D(t)}{NR^2_0(t)} \right )^{-1}
\end{equation}
Finally $\beta(t+1)$ and $v_1(t+1)$ are found by
\begin{equation} \mylabel{A12}
\beta_i(t+1)=m(t)\left(\beta_i(t)+\frac{ D_i(t)}{R_0(t)}\right)
\end{equation}
Inserting $R_k(t)$ leads to the formula given in section \ref{multi}.

Necessary conditions on the GARCH parameters $\alpha_{k}$ and $\gamma_{k}$
are obtained by setting $r_{Mt}^2=\bar{v}_{0}$ and $r^2(t)=N(\bar{v}_{0}+\bar{v}_{1})$ to its
mean values. Then the recursions have a fixed point $v_k(t)=\bar{v}_k$ and
$\beta(t)=\bar{\beta}$ which is stable for
\begin{equation} \mylabel{A13}
0<\gamma_{k}<\gamma_{k}+\alpha_{k}<1
\end{equation}

\section{Calculation of the likelihood  \label{like}}
The log likelihood with noise distribution $f(\varepsilon)$ can be computed using
\begin{equation} \mylabel{A18}
\varepsilon(t)={H(t)}^{-1/2}  r(t)
\end{equation}
With a spectral decomposition for $H(t)$ functions of $H$ can be calculated analytically.
For $H(t)^{-1/2}$ we obtain
\begin{equation} \mylabel{A19}
H(t)^{-1/2}=\frac{1}{\sqrt{Nv_0(t)}}\; P_0(t) \;+\frac{1}{\sqrt{v_1(t)}}\;P_1(t)
\end{equation}
It is easy to verify $(H(t)^{-1/2})^2\cdot H(t)=1$ with the orthogonality relations
for $P_\nu(t)$. For $\varepsilon$ we obtain
\begin{equation} \mylabel{A16}
\varepsilon(t)=\frac{r_{M}(t)}{\sqrt{Nv_0(t)}}\; \beta(t) \;+\; 
      \frac{1}{\sqrt{v_1(t)}}\; (r(t)-r_{M}(t)\beta(t))
\end{equation}
The log likelihood $L$ is given by
\begin{equation} \mylabel{A17}
L=\frac{1}{2}\sum_t\left[\sum_i \ln f(\varepsilon_{i}(t))-\ln(Nv_0(t)) -(N-1)\ln v_1(t)\right]
\end{equation}
For Gaussian noise calculation of the $\ln(f)$ can be avoided. The complication
of using t-distributed noise leads to an negligible increase of computing time
compared to the calculation of $v_\nu(t)$ and $\beta(t)$. In any case the
computing time increases only with $T \cdot N$.

\section{Covariance targeting for the restricted covariance matrix  \label{noise}}

For the time averaged covariance matrix $\tilde{C}$ of our model computed from
\begin{equation} \mylabel{c1}
r(t)=\bar{H}^{1/2}  \varepsilon(t)
\end{equation}
we decompose the noise $\varepsilon$ into a component $\varepsilon_M(t)$ parallel to 
$\bar{\beta}$ and a component $\bar{\varepsilon}_i(t)$ perpendicular to
$\bar{\beta}$. Both are i.i.d. with mean zero and variance one for large $N$. For
$\tilde{C}$ we obtain
\begin{eqnarray} \mylabel{c2}
\tilde{C} &=&\frac{1}{T} \sum_t \left[ N\;v_0(t) \;P_0(t)\; \varepsilon_M^2(t) + 
    v_1(t) \; \bar{\varepsilon}(t)  \bar{\varepsilon}'(t) \right. \nonumber \\
  &+& \left. \sqrt{v_0(t)v_1(t)}\; \varepsilon_M(t)(\beta(t)  \bar{\varepsilon}'(t)+
      \bar{\varepsilon}(t) \beta'(t)) \right ]
\end{eqnarray}
The statistical uncertainty of $\tilde{C}$ is mainly due to the noise and much less due
to the fluctuations of the slowly varying $v_\nu(t)$ and $\beta(t)$. For a rough estimate
we neglect in the time average the $\varepsilon$-dependence in the latter. For the
$tr(\tilde{C})$ we get for large $N$
\begin{equation} \mylabel{c2a}
tr(\tilde{C})=\frac{1}{T} \sum_t\left[ N\;v_0(t) \;\varepsilon_M^2(t) + v_1(t) \;
               \sum_i\; \bar{\varepsilon}_i^2(t) \right ]
\end{equation}
The law of large numbers leads for the average over the noise to
\begin{equation} \mylabel{c2b}
tr(\tilde{C})=\frac{N}{T} \sum_t \; (v_0(t) \; + v_1(t))=N(\bar{v}_0+\bar{v}_1)
\end{equation}
where we replaced the time average of $v_\nu(t)$ by its mean $\bar{v}_\nu$. 
Comparing equation \myref{c2b} with the observed covariance matrix gives one relation for
 $\bar{v}_\nu$ with a relative error of $1/\sqrt{T}$. The same procedure leads for
the third term in equation \myref{c2} to a contribution of order $1/\sqrt{T}$ since the 
expectation value of $\bar{\varepsilon}_i(t)\; \varepsilon_M(t)$ vanishes. Replacing 
the second term $ v_1(t)$ by $\bar{v_1}$ leads to a random matrix $R$ with a 
Mar\v{c}enko-Pastur-spectrum. The average of the first term in 
equation \myref{c2} is equal to $N\bar{v}_0\bar{P}_0$. The resulting $\tilde{C}$ reads as
\begin{equation} \mylabel{c5}
\tilde{C}=N\;\bar{v}_0 \;\bar{P}_0\;+\; R
\end{equation}
Since $|R_{ij}|<<N$ we can determine $\bar{v}_0$ and $\bar{\beta}$ by comparing the leading
eigenvalue and eigenvector of $\tilde{C}$ and the observed $C$. The statistical 
error is in the order of $1/\sqrt{T}$.

\section{\label{sec:chi} Experimental distributions}

For comparing empirical data with simulations we apply a qualitative criterion 
which allows to locate eventual disagreement and is less sensitive to systematic
errors as outliers.
We assume $T$ independent measurements of an observable $x(t)\; (t=1,T)$. The values
$x(t)$ are binned into $M$ bins of width $\Delta x_m$ centered around $x_m$ containing $n_m$
events. The probability density $f_m$ at $x_m$ is given by the relative frequency
\begin{equation} \mylabel{d1}
f_m=\frac{1}{\Delta x_m}\;\frac{n_m}{T}
\end{equation}
Errors of $f_m$ can be estimated assuming a Gaussian distribution by
\begin{equation} \mylabel{d2}
\Delta f_m=\frac{1}{\sqrt{n_m-1}}\;f_m
\end{equation}
Good plots of the distribution can be achieved by varying $\Delta x_m$ and by using a minimum  for $n_m \ge 6$. In this way the empirical distributions in figures \ref{c_spectrum},
\ref{degarch}, and \ref{pdf_chisq} were created.

To estimate the quality of agreement of $f_m$ with a simulated density $F(x)$
we adopt the following qualitative measure: the average mean squared deviation 
also called $\chi^2$ ratio
\begin{equation} \mylabel{d3}
\chi^2=\frac{1}{M}\sum_{m=1}^M\; \left(\frac{F(x_m)-f_m}{\Delta f_m}\right)^2
\end{equation}
A value of $\chi^2<4$ can be interpreted as agreement with 5\% confidence level and
values of $\chi^2<10$ still signal qualitative agreement.

\section{Distributions of covariances \label{sec:appcov}}

\begin{figure}[htb]
\begin{center}
\includegraphics[width=0.8\linewidth,trim= 60 10 60 10, clip=true]{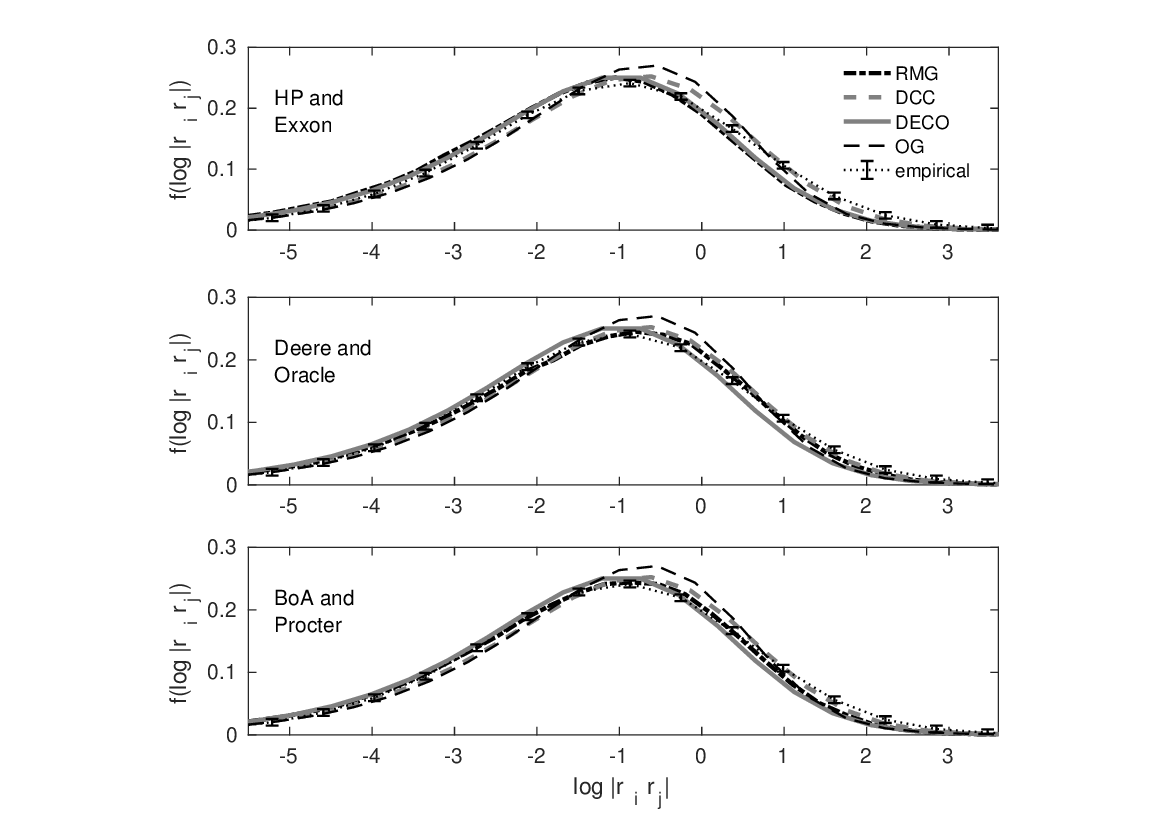}
\end{center}
\caption{\em Simulated distributions of covariances. We show the distributions of the
        predicted covariances versus the original data. The dotted line connects the
        empirical pdf and a two standard deviations range. The broken and dotted black 
        line corresponds to the simulated predicted correlations of RMG. The results for 
        the DCC model are represented by the broken gray line The DECO model is represented 
        by the gray line and the OGarch model by the broken black line. Note that 
        the horizontal axis shows the logarithm of the absolute values of $r_i r_j$, 
        therefore cases where the correlation is 0 are omitted.
        }\label{fig:5dist}
\end{figure}

For a visual comparison of the distributions we use the variable $\log(|r_i r_j|)$ for \mbox{$|r_i r_j| > 0$} in order to be sensitive to large values of $r_i r_j$ and not to be overwhelmed by the
many small $r_i r_j$. Three typical examples of these distributions for pairs of stocks are 
shown in figure \ref{fig:5dist}. The results for OG are often outside the empirical errors,
whereas RMG and DECO agree with DCC and the data. 

\end{appendix}

\end{document}